# Adaptive and robust smartphone-based step detection in multiple sclerosis

Lorenza Angelini,[1] Dimitar Stanev,[1] Marta Płonka,[2] Rafał Klimas,[1] Natan Napiórkowski,[2] Gabriela González Chan,[3] Lisa Bunn,[3] Paul S Glazier,[3] Richard Hosking,[4] Jenny Freeman,[3] Jeremy Hobart,[5] Jonathan Marsden,[3] Licinio Craveiro,[1] Mike D Rinderknecht,[1*] and Mattia Zanon[1*]

* Shared last author

[1]F. Hoffmann-La Roche Ltd, Basel, Switzerland

[2]Roche Polska Sp. z o.o., Warsaw, Poland

[3]School of Health Professions, University of Plymouth, Plymouth, UK

[4]University of Plymouth Enterprise Ltd, Plymouth, UK

[5]Peninsula Medical School, Faculty of Health, University of Plymouth, Plymouth, UK

**Corresponding author**

Mike D Rinderknecht

mike.rinderknecht@roche.com

# Abstract

**Background:** Many attempts to validate gait pipelines, a framework that processes sensor data to detect gait events for subsequent computation of gait-related measures, have focused on validating the detection of initial over final contacts in supervised settings using a single sensor.

**Objective:** To evaluate the performance of a gait pipeline in detecting initial and final contacts, (thus enabling computation of a larger set of gait measures) using a step detection algorithm adaptive to different environments, smartphone wear locations, and gait impairment levels**.**

**Methods:** In GaitLab, healthy controls (HC) and people with multiple sclerosis (PwMS; Expanded Disability Status Scale 0.0–6.5) performed supervised laboratory-based walking activities (structured in-lab overground and treadmill Two-Minute Walk Test [2MWT]) during two on-site visits carrying six smartphones (left/right front trouser pocket, front/back belt and left/right back trouser pocket) and unsupervised walking activities (structured and unstructured real-world walking) once daily for 10–14 days using a single smartphone. Reference gait data were simultaneously collected with either a motion capture system (structured in-lab treadmill 2MWT) or Gait Up inertial measurement unit sensors (all other assessments). The gait pipeline's performance in detecting gait events (initial/final contacts) was evaluated through F1 scores (harmonic mean of precision and recall; tolerance window: 0.5s) and absolute temporal error with respect to reference measurement systems.

**Results:** We studied 35 HC and 93 PwMS. Gait events were accurately detected across all smartphone wear locations. Median F1 scores for initial/final contacts on in-lab walking activities were ≥99.0%/≥97.6% in HC and ≥99.0%/≥98.2% in PwMS. Median (IQR) F1 scores remained high on structured real-world walking (HC: 100% [99.9%–100%]/100% [99.8%–100%]; PwMS: 99.9% [99.0%–100%]/99.6% [98.2%–100%]) and unstructured real-world walking (HC: 97.8% [96.2%–98.8%]/97.8% [95.8%–98.6%]; PwMS: 94.4% [90.4%–96.6%]/94.0% [90.0%–96.5%]). Median temporal errors for both initial/final contacts were ≤0.08s. Neither age, sex, disease severity, walking aid use, nor environment (indoors/outdoors) impacted significantly on pipeline performance (all $p>0.05$).

**Conclusion:** This gait pipeline accurately and consistently detected initial and final contacts in HC and PwMS across different smartphone wear locations and environments, expanding previous research and highlighting its potential for real-world gait assessment.

**Trial registration:** ISRCTN15993728

# Introduction

Gait impairment is commonly observed among people with multiple sclerosis (PwMS) [1, 2]. Slowing or stopping worsening of walking disability can have a profound impact on affected patients and the burden the disease has on society [3-6]. Different scales and performance measures are available to assess and monitor gait impairment including the Two-Minute Walk Test (2MWT) [7], Timed 25-Foot Walk (T25FW) [8], Timed Up and Go (TUG) [9], and Expanded Disability Status Scale (EDSS) [10]. Despite their use, they have significant limitations in capturing an individual's functional abilities in daily life and identifying subtle changes in gait, particularly in the early stages of disease. These limitations arise from the requirement for a controlled, supervised setting and the presence of a trained examiner, which differs from the challenges that PwMS face daily [11]. Furthermore, these scales and performance measures only provide a single value, whether it is the time needed to complete the task or a score, and thus only partially reflects the complex and dynamic nature of mobility, such as the capacity to navigate uneven surfaces or respond to unexpected obstacles [12]. Alternative assessment methods, such as wearable technologies, may offer a comprehensive and quantitative approach, providing more detailed information on several gait-related characteristics in daily life [13-19]. Smartphones, in particular, have emerged as a promising tool for frequent, cost-efficient, convenient, and minimally intrusive assessments in a home environment, offering a much larger set of measures to probe different aspects of functional ability [20-22]. Incorporating these alternative assessment methods ensures a more accurate assessment of functional impairment, enabling the development of more effective therapeutic options for PwMS.

Multiple pipelines have been developed in recent years to extract digital outcomes from gait data recorded using inertial measurement units (IMUs) embedded in smartphones or wearable devices [13, 23-26]. However, variations in sensor placement, disease characteristic, real-world environments such as indoor versus outdoor environments (also referred to as contextual factors [27]), or use of walking aids can significantly impact the technical validity of step detection algorithms integrated in gait pipelines (i.e., the extent to which such algorithms can accurately and reliably detect gait events when compared against reference measurement systems), especially in populations with pathological gait, including PwMS.

Many validation studies have predominantly used IMUs with a fixed wear location, such as the lower back or the ankles [14, 16, 28]. And while some gait pipelines are available that are robust against different sensor placements, they are typically designed to count steps and are, therefore, restricted to detecting initial contacts only, thus limiting the range of digital gait measures they can compute [29, 30]. Furthermore, technical validation studies have been mainly conducted in controlled, supervised settings, such as laboratories or clinics, which do not reflect performance in everyday life [31-33]. Such studies have also mostly focused on healthy controls (HC) [29, 34, 35], with only limited studies addressing populations with pathological gait [36, 37].

These limitations highlight the importance of conducting validation studies that examine the potential impact of both different smartphone wear locations that reflect how smartphones are carried in everyday life and different environments and settings on the detection of initial and final contacts (i.e., heel strike and toe-off events, respectively) in a representative sample of individuals with relevant pathologies. Addressing these issues is critical for developing a robust gait pipeline that can reliably characterize gait abnormalities in daily life contexts across diverse patient populations.

The GaitLab study (ISRCTN15993728) was designed to collect data for the development and validation of a gait pipeline that is robust with regards to different intrinsic gait factors (e.g., population type, disease stage, walking speeds), test conditions (supervised, in-lab vs unsupervised, real-world; structured vs unstructured walking), real-world environments (indoor vs outdoor environments), smartphone wear locations, and the use of walking aids and orthotics [38]. Using data from this study collected from HC and PwMS, we establish here the technical validity of a gait pipeline, which includes a novel step detection algorithm adaptive to different test conditions, real-world environments, smartphone wear locations, and gait impairment levels. In particular, we describe the algorithm's performance in accurately identifying the gait events within a gait cycle such as initial and final contacts and evaluate the impact of intrinsic gait factors and other potential confounding factors on the algorithm’s performance.

# Methods

## Study design and participants

The GaitLab study employed a well-defined experimental protocol and state-of-the-art equipment to develop and validate the gait pipeline. The study design has been previously described in detail [38]. In short, the study recruited PwMS and age- and sex-matched HC from a single site in Plymouth, United Kingdom, from January 2022 to October 2023. All participants were required to be at least 18 years old. PwMS were additionally required to have a confirmed diagnosis of either relapsing-remitting, primary progressive, or secondary progressive MS, per the 2010 or 2017 McDonald diagnostic criteria [39, 40]; the ability to walk for a period of at least 6 minutes; and a baseline EDSS score between 0.0 and 6.5, inclusive. The protocol consisted of two visits to a gait laboratory and a two-week period of unsupervised remote testing in the home environment in between (Figure 1). The technical validity of the gait pipeline and, in particular, of the novel and adaptive step detection algorithm integrated in the pipeline was evaluated using gait data obtained during the second on-site visit as well as during the remote testing period. Study participants were allowed to use their usual walking aid and or orthotic if needed, and rest as required.

### In-lab walking activities

During the second on-site visit, each participant was requested to perform a structured in-lab overground walking 2MWT and a structured in-lab treadmill walking 2MWT (Figure 1). For the first activity, participants were instructed to complete a 2MWT, which involved walking back and forth along a marked 10-meter corridor, while walking safely at their fastest possible pace. For the second activity, participants completed the 2MWT on a dual-belt treadmill (Motek,

Netherlands), which featured a self-paced algorithm to enable participants to walk at a self-selected fast pace, i.e., a pace that resembles overground walking most closely [41]. During the treadmill session, participants could hold the handrails in the event of losing balance and were outfitted with a safety harness to prevent falls.

Participants completed both activities while carrying six smartphones with a built-in IMU (Samsung Galaxy A40 running on Android version 9; sampling frequency: 50 Hz; weight: 140 g) in customized shorts and adjustable waist belts to capture gait data. Each smartphone was placed in different pockets and belts. These were the left/right front trouser pocket, belt back, belt front, and left/right back trouser pocket (Figure 1).

## Real-world walking activities

During the unsupervised remote testing period conducted in the home environment, participants performed unsupervised and self-administered structured and unstructured real-world walking activities in real-life conditions for 10 – 14 days. The structured real-world walking activity entailed a daily 2MWT, during which participants were directed to walk as fast and far as possible but safely for 2 minutes while minimizing sharp turns as much as possible, mimicking the clinician-administered 2MWT. This constraint was introduced to ensure that this activity closely resembles the clinician-administered 2MWT. Participants were required to carry a single smartphone, which was preferably placed in a waist-worn front belt bag (Figure 1). The structured real-world walking activity could be conducted indoors or outdoors, with a preference for alternating between the two on consecutive days if possible. This walking activity involved the study participants carrying a single smartphone for up to 4 hours a day, with at least 15 minutes of walking while they went about their typical daily routine. Study participants were allowed to position the smartphone in their preferred location, ensuring it remained as close to their body as possible (Figure 1).

Following each structured and unstructured real-world walking activity, participants were required to update their daily gait diary, in which they captured information about how they performed the gait tests remotely, such as the smartphone's wear location, whether the walking activity was performed indoors or outdoors, the terrain (e.g., flat versus hilly), or walking aids if used. The gait diary was designed to provide additional context for the interpretation of the collected gait data.

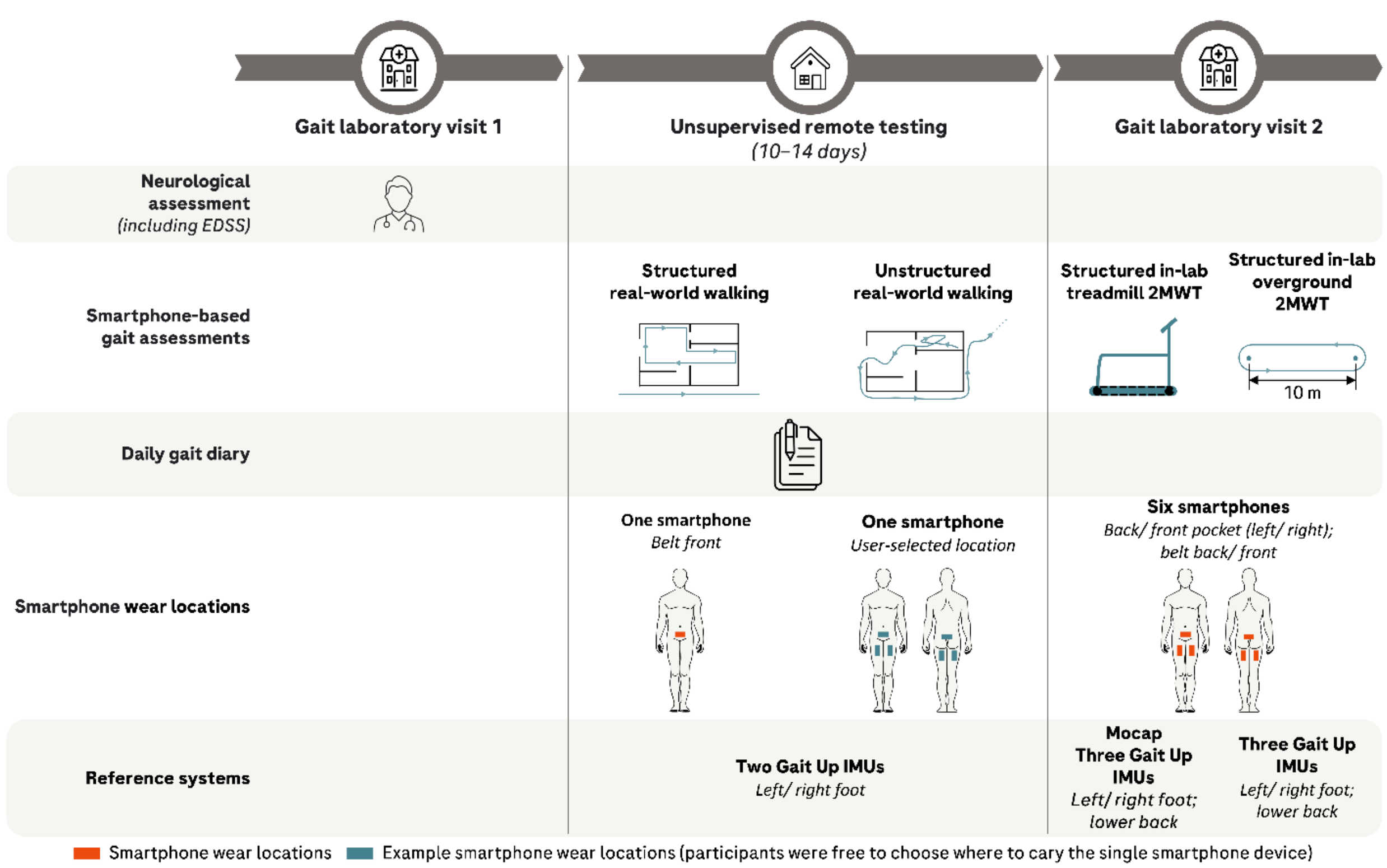

**Figure 1. Outline of the clinical visits and relevant data used in the analysis.** The gait pipeline was validated using gait data collected from the structured real-world walking, unstructured real-world walking, structured in-lab treadmill 2MWT, and structured in-lab overground 2MWT. Data collected during the neurological assessment and with the daily gait diary were used to provide additional context for the interpretation of the gait pipeline's performance metrics. A full overview of all assessments and data collected in the study is reported in [38]. 2MWT = Two-Minute Walk Test; EDSS = Expanded Disability Status Scale; IMUs = inertial measurement units.

## Reference measurement systems

To evaluate the step detection algorithm's technical validity, ground-truth data from two reference measurement systems were collected simultaneously with smartphone data. The reference measurement systems were synchronized to the smartphones as described in section "Supplementary appendix" and in [42]. The "gold standard" reference system consisted of a marker-based motion capture (mocap) system and two multiaxial force plates. The mocap system included 12 infrared cameras (Vicon Vero™ v1.3 cameras) with a resolution of 1.3 MP at 100 frames per second and 26 reflective markers (14 mm in diameter) positioned on body landmarks based on the Motek Human Body Model marker set [43]. Marker trajectories used to capture participant motion were recorded at a sampling frequency of 100 Hz. The two multiaxial force plates (ForceLink R-Mill force plates, Motek, Netherlands), embedded within the treadmill, enabled precise readings of the ground reaction forces (GRFs) and moments at a sampling frequency of 1000 Hz. This gold standard reference system was used in the gait laboratory during the structured in-lab treadmill 2MWT.

The "silver standard" reference system consisted of body-worn triaxial accelerometer and gyroscope sensors (i.e., IMUs) manufactured by Gait Up (Physilog 6® model, Gait Up, Lausanne Switzerland). The Gait Up IMUs recorded continuous 3D accelerometer and gyroscope data at 128 Hz. Two configurations of the Gait Up system were deployed: (I) a three-sensor setup (one IMU on each foot and an additional IMU at the lower back near the center of mass) used during the structured in-lab overground and treadmill 2MWTs, and (II) a two-sensor setup, with one IMU on each foot, used during the structured and unstructured real-world walking activities.

## Gait pipeline

A gait pipeline in biomechanics and movement analysis represents a systematic framework for evaluating human mobility [44]. It consists of an algorithmic and processing pipeline that segments raw IMU sensor signals into gait bouts, or walking segments, and detects gait events (i.e., initial and final contact), which can be used for subsequent computation of digital gait measures that measure the quantity and quality of gait. Specifically, the gait pipeline introduced here comprises the following key steps: preprocessing, alignment with gravity, task recognition, alignment with the anatomical reference frame, adaptive step detection, and computation of digital gait measures assessing the quality and quantity of gait (Figure 2). However, the evaluation of digital gait measures goes beyond the scope of the presented analysis, which focuses on the performance of the adaptive step detection algorithm in detecting initial and final contacts.

### Signal preprocessing

Initially, the data collected from the accelerometer and gyroscope were resampled to the same time grid and temporally aligned. Following this, the accelerometer's raw signal was filtered using a low-pass filter (second-order double-pass Butterworth filter with a cutoff frequency of 17 Hz [45]).

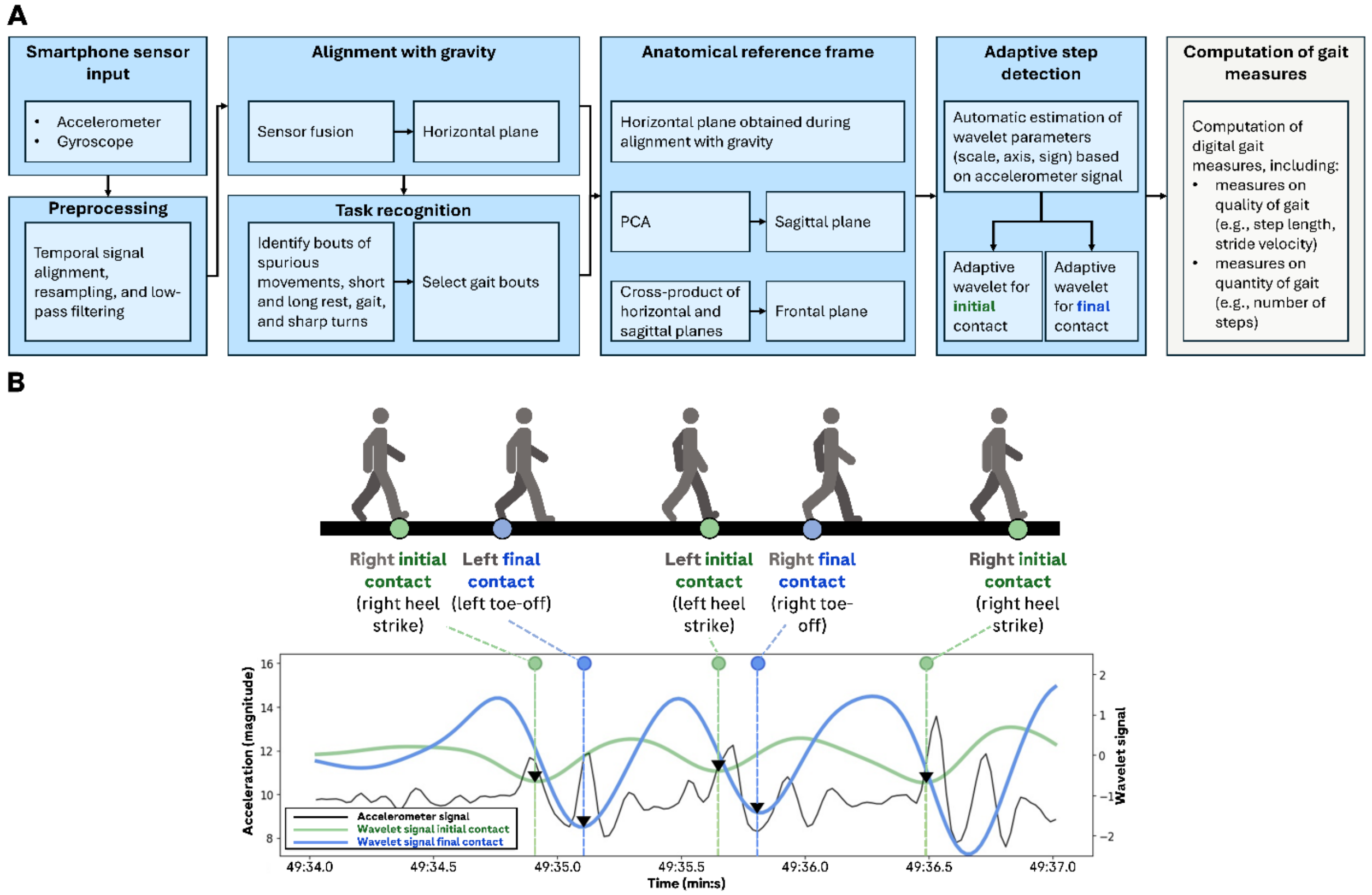

**Figure 2. Gait pipeline.** (A) The gait pipeline consists of six major steps: preprocessing, alignment with gravity, task recognition, alignment with the anatomical reference frame, adaptive step detection, and calculation of gait measures. The preprocessing, alignment with gravity, and alignment with the anatomical reference frame steps aim to smooth and align the accelerometer and gyroscope signals in both time and space. The task recognition step aims to identify bouts of spurious movements, rest periods, sharp turns (i.e., turns of >90 degrees), and gait bouts without sharp turns that last at least 2 seconds. Here, adaptive step detection was subsequently applied to gait bouts (the step detection algorithm was not applied to sharp turns, as the reference measurement systems did not robustly detect steps taken during sharp turns). (B) In the adaptive step detection step, parameters of the continuous wavelet transform, such as wavelet scale, wavelet axis, and wavelet sign, are automatically estimated from the accelerometer signal. Two separate wavelet transforms are applied along the axis established by the *wavelet axis* and at the scale specified by the *wavelet scale*, resulting in two wavelet signals: one represented by a green curve for detecting initial contacts (green dots) and another represented by a blue curve for detecting final contacts (blue dots). These gait events are identified in the respective wavelet transform as local minima or maxima, depending on the *wavelet sign* (here: local minima). PCA = principal component analysis.

## Alignment with gravity

The direction of gravity in the preprocessed accelerometer and gyroscope signal, and hence the horizontal plane (Supplementary Figure S1), was determined through sensor fusion using an orientation estimation algorithm (Madgwick filter with gain parameter set to $\beta = 0.041$ [46]). The magnetometer signal was deliberately not used in this step due to potential magnetic interference, in particular in indoor walking scenarios.

## Task recognition

Having the smartphone accelerometer and gyroscope data aligned with the direction of gravity enabled the identification of bouts of spurious movements, short and long rests, sharp turns, and gait (or walking).

Moving (i.e., bouts of spurious movements, sharp turns, and gait) and non-moving (i.e., resting) intervals were identified using both acceleration and gyroscope signals. The process involved splitting the acceleration and gyroscope signals into non-overlapping intervals. The default window length was set to 0.6 s, which aligned with the typical mean step duration among PwMS. However, it can be adjusted based on the specific population of interest.

To identify non-moving intervals, three criteria needed to be satisfied within an interval. First, the accelerometer magnitude must fall within a defined range, specified as a reference value ± 10% of the reference value. Here, a reference value corresponding to the gravitational acceleration of $g = 9.81\ m/s^2$ was used. Second, the threshold for gyroscope magnitude should ideally lie between 0.2 rad/s and 0.6 rad/s, where lower values more rigorously filter non-moving windows. A 0.2 rad/s tolerance indicates static balance as a non-moving interval, while 0.6 rad/s ensures adequate handling of minor movements across different smartphone wear locations. Here, a threshold of 0.6 rad/s was used. Finally, the combined standard deviation of the accelerometer signal should remain below $0.2\ m/s^2$. This threshold is effective for general performance as a value of $0.05\ m/s^2$ identifies static balance and a value of $0.4\ m/s^2$ allows for some movement tolerance. All three thresholds were derived empirically from moving and non-moving data recorded during a dry run prior to the study.

Non-moving intervals less than 1 second apart were merged into a single non-moving interval. Subsequently, non-moving intervals were classified as rest if surrounded by moving intervals, or as boundaries if they were within 2 seconds of the start or end of the recorded data.

Intervals that were not categorized as non-moving intervals were labelled as moving intervals. Specifically, moving intervals shorter than 2 seconds were deemed unknown, as they might include spurious movements, and were discarded in subsequent processing. The threshold of minimum gait bout duration was set at 2 seconds based on the definition of a gait bout in Kluge et al. [47], as a sequence containing at least two consecutive strides (= 4 steps). Hence, with a mean step duration of 0.5 seconds and a minimum of four steps per gait bout, the minimum gait bout duration is 2 seconds. Hence, moving intervals longer than 2 seconds were marked as possible gait bouts. After establishing the start and end points of these moving intervals (i.e., gait bouts), the segmented data underwent further analysis and step detection was applied to detect gait events (see section “Adaptive step detection” below).

Step detection was only applied to the gait bouts that did not contain a rest interval nor sharp turns. Periods of turns were intentionally excluded as steps during sharp turns could not be consistently detected with reference measurement systems. The detection of sharp turns relied on the gyroscope signal, utilizing a well-established and commonly adopted algorithm [48]. Here, only sharp turns of ≥ 90 degrees were considered.

### Verification of the moving interval as a gait segment

Autocorrelation is fundamentally the correlation of a signal with a time-shifted version of itself, making it a useful tool for identifying periodic patterns in data. As gait produces a periodical pattern in smartphone sensor data, bouts initially identified as gait were verified as true gait bouts by an autocorrelation algorithm [44].

## Alignment with the anatomical reference frame

For the step detection algorithm to be agnostic to the smartphone’s orientation, the smartphone sensor data must be transformed into an orientation-agnostic anatomical reference frame (Supplementary Figure S1) [44]. The reference frame's horizontal plane was already obtained during alignment with gravity. The other two planes were identified as follows. The sagittal plane, which is parallel to the direction of movement, was determined using principal component analysis (PCA) applied to the gravity-aligned accelerometer data. This method relies on the fact that the direction of movement corresponds to the maximum variance in the data (i.e., maximum movement). Finally, the frontal plane was established by computing the cross product of the horizontal and sagittal planes. As PCA requires a sufficient amount of data—at least 3 to 4 seconds of samples—to yield reliable results [49-51], the accuracy of the identified sagittal and frontal planes was verified through an autocorrelation-based algorithm (see section "Verification of sagittal and frontal planes through autocorrelation" in Supplementary appendix) [44].

The normal vectors of these three planes were subsequently used to build the transformation matrix that defined the orientation-agnostic global coordinate system. Following this transformation, the vertical axis of the smartphone sensor data was realigned with the gravitational vector and pointing upward; the antero-posterior axis was realigned with the direction of movement, as determined by the PCA, and pointing either forward or backward. The medio-lateral axis was then derived as the cross-product of the newly realigned vertical and antero-posterior axes, thus pointing to the left or to the right accordingly.

## Adaptive step detection

The continuous wavelet transform ('CWT'; A mathematical function resembling a tiny wave used to analyze and capture both high and low-frequency details in data, helping break down complex patterns into simpler component) was used to identify initial contact and final contact events within the gait cycle [52]. The wavelet scale, wavelet axis (the axis along which the wavelet function is applied) and wavelet sign were automatically determined based on the input acceleration and gyroscope signals [44]. This approach enhances the robustness of the pipeline by making the step detection adaptive to the characteristics of the input signals. Once these parameters were selected, the CWT (PyWavelets v1.8.0 package [53]) was applied for initial and final contact detection as follows:First, the acceleration along the wavelet axis was integrated and differentiated using a Gaussian CWT, with initial contacts detected as the timepoints of the minima or maxima, depending on the wavelet sign.Next, the differentiated signal underwent further CWT differentiation, from which final contacts were identified as the time points of the minima or maxima based on the wavelet sign. Finally, the laterality of each initial contact was determined using the sign of the gyroscope signal.

A quality check of gait events was performed to remove false detections based on physiological limits (i.e., gait events occurring too early or too late were removed) and to verify whether the event sequence matched the standard gait cycle (from initial contact with one leg until the next initial contact on the same leg; see also top panel in Figure 2B). Furthermore, any final contacts beyond 25% of the maximum estimated stride duration were excluded. At the 25% mark of the gait cycle, the limb that initiated contact reaches the midstance phase, indicating that toe-off has already happened [54]. The rationale for not examining lower percentages is to ensure sufficient tolerance, thereby reducing the risk of missing any valid final contacts that may occur. Maximum stride duration was estimated from acceleration or gyroscope signals using an autocorrelation-based algorithm described in [44]. Specifically, this algorithm uses autocorrelation to estimate the step duration. As strides consist of two steps, stride duration could be derived by simply multiplying the estimated step duration by 2. However, an additional 50% tolerance was applied to compute the maximum estimated stride duration (i.e., maximum estimated stride duration = 3 * estimated step duration).

## Statistical analysis

The analyses presented here evaluate the gait pipeline's performance in detecting gait events such as initial and final contacts. These gait events can subsequently be used to compute spatial, spatiotemporal, and temporal gait measures, such as step/stride length, stride/step duration, and step/stride velocity, as well as measures from other gait domains (e.g., measures related to gait asymmetry and variability). The evaluation of the psychometric properties of these gait measures, however, goes beyond the scope of the analysis presented in this publication and will be reported in dedicated publications.

### Gait pipeline's performance and temporal precision of gait event detection

The technical validity of the gait pipeline was evaluated by assessing the accuracy and temporal precision in identifying gait events. Detection performance and temporal precision were computed separately for the following groups: HC, all PwMS, PwMS with mild EDSS scores (EDSS ≤ 3.5), PwMS with moderate EDSS scores (EDSS = 4.0 – 5.0), and PwMS with severe EDSS scores (EDSS = 5.5 – 6.5). Gait events detected in smartphone sensor data were compared to those detected in reference measurement system data (mocap and Gait Up data, see section "Supplementary appendix") within a temporal tolerance window of 0.5 s, centered around the event identified by the reference tool. The 0.5-second tolerance window was selected based on its representation of a step duration and agreement with previous research [26]. Furthermore, this window also allows for some robustness in case the smartphones and reference measurement systems are not perfectly synchronized with each other, especially during structured and unstructured real-world walking, where participants were responsible for synchronizing the devices.

Gait events were classified as true positives (TPs) if identified by both the smartphone and the reference system; false positives (FPs) if detected by the smartphone but not by the reference system; or false negatives (FNs) if missed by the smartphone but identified by the reference system. If the smartphone found multiple events within the tolerance window, the event closest to the reference event was labelled a TP, while the others were marked as FPs, assuming they did not correspond to any subsequent reference event. Following the categorization of events as TPs, FNs, or FPs, the algorithm's performance in detecting initial and final contacts was evaluated using the following performance metrics:

$$Precision = \frac{TP}{(TP + FP)}$$

$$Recall = \frac{TP}{(TP + FN)}$$

$$F1\ score = 2\frac{(Precision\ \times Recall)}{(Precision + Recall)}$$

Precision measures the proportion of TPs among all events detected by the smartphone, while recall, also known as sensitivity, indicates the proportion of TPs among all events that should have been detected. The F1 score was computed as the harmonic mean of precision and recall, providing a balanced measure of the algorithm's performance [55].

Additionally, temporal error metrics were computed to measure the temporal difference between smartphone and reference gait events (i.e., this difference is positive if the smartphone gait event occurs after the reference gait event and negative if it happens earlier). These metrics were calculated across all steps detected by the smartphone during a single test (i.e., all TP gait events) and included the constant error (mean error across steps), absolute error (mean of absolute errors across steps [i.e., bias]), variable error (standard deviation of errors across steps [i.e., consistency]), total variability mean (root mean square divided by the number of steps), and interquartile range (IQR) error.

Performance and error metrics were aggregated as follows: For structured in-lab treadmill and structured in-lab overground 2MWTs—each performed once per participant—, performance metrics were summarised across participants using the median, IQR, first quartile (Q1), third quartile (Q3), 5th percentile, 95th percentile, mean, and 95% confidence interval (CI). Error metrics were aggregated by computing the mean and 95% CI across participants. In contrast, for structured and unstructured real-world walking activities, the aggregation involved two steps, as each participant completed these activities multiple times during remote testing. First, within each participant, the median and IQR were calculated across tests and days for both performance and error metrics, resulting in a single value per individual (i.e. the within-participant median and IQR). Next, the within-participant median and IQR were pooled across participants by computing the median, IQR, Q1, Q3, 5th percentile, 95th percentile, mean, and 95% CI for performance metrics, and the mean and 95% CI for error metrics. Additionally, to quantify the average of within-participant performance variability at the group level, the median of the within-participant IQR was computed for the performance metrics, referred to as the 'within-subjects IQR' (ws-IQR).

## Effect of potential confounding factors on gait event detection

As confounding factors such as age, sex, or MS disease status can impact gait [56, 57], a Bayesian generalized linear model [58] was used to evaluate whether such confounding factors affected the performance of the step detection algorithm (i.e., the F1 scores):

$$logit(\mu[i]) = \alpha + \beta \times \overline{age} + s(sex_{idx}[i]) + d \times \sum_{j=0}^{disease_{idx}[i]-1} \delta_j + u(subject_{idx}[i]) + e(environment_{idx}[i]) + h(aid_{idx}[i])$$

Predictor variables, or fixed effects, included the global intercept (*α)*, subject intercept to account for repeated measures within subjects (*u*), age (*β*), sex (*s*), total disease effect (i.e., both disease status and disease severity (*δ))*, environment (here: indoors vs outdoors) (*e*), and walking aid usage (*h*). Parameter *δ*, which indicates an increment between disease status and severity, was used to constrain the effect of disease as an ordinal variable. The term "*idx*" signifies index variables. Further details on the mathematical definition of the Bayesian model can be found in the section "Supplementary appendix". This analysis was conducted on data collected from the structured real-world walking activity.

# Results

## Baseline demographics and disease characteristics

A total of 35 HC and 100 PwMS were enrolled, of whom 35 HC and 96 PwMS had smartphone data available. A small subset of participants was excluded from the analyses due to malfunctioning reference systems, failure to use the devices during the assessments, synchronization issues between the smartphones and reference systems, incorrect identification of gait events by the reference systems, and missing or incorrect test executions. Thus, smartphone and reference measurement data were available for 32 – 35 HC and 79 – 87 PwMS across the different test conditions (Figure 3). Overall, HC and PwMS contributed 68 – 70 minutes and 158 – 172 minutes of in-lab gait data, respectively (Figure 3). More gait data was collected in the real-world setting as participants were instructed to complete the walking activities multiple times (i.e., once daily) rather than just once. HC provided 548 minutes of gait data during structured real-world walking and 729 minutes during unstructured real-world walking. PwMS provided 1422 minutes of gait data during structured real-world walking and 1237 minutes during unstructured real-world walking, respectively (Figure 3).

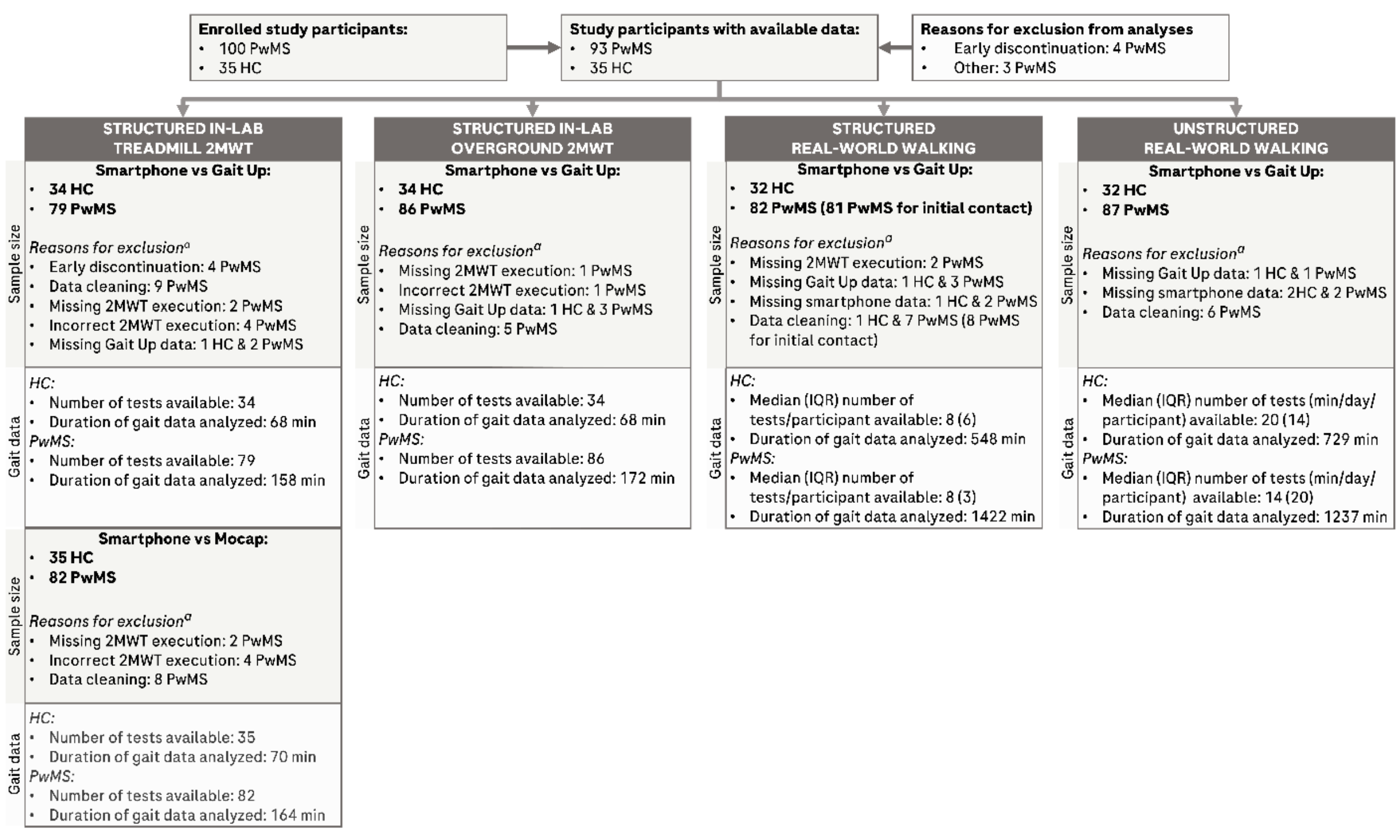

**Figure 3. Number of participants with available smartphone and reference measurement systems data and available gait data for each of the different test conditions.** Individual data were excluded if they did not meet the quality checks due to reference system malfunction, synchronization failure, incorrect identification of gait events by reference, and missing or incorrect test executions. [a]Considering all 35 HC and 100 PwMS enrolled in the study. 2MWT = Two-Minute Walk Test; HC = healthy controls; PwMS = people with multiple sclerosis.

The baseline demographics and disease characteristics are reported in Table 1. HC and PwMS were similar in age, height, weight, and body mass index. PwMS were predominantly diagnosed with relapsing-remitting disease (70% of PwMS), and the median (IQR) EDSS score among all PwMS was 5.0 (4.0 – 6.0). Furthermore, the PwMS cohort represented a wide range of gait impairment: T25FW times ranged between 2.9 s and 18.6 s (median [IQR]: 5.4 s [4.5 s – 8.2s]), and MSWS-12 transformed total scores between 0.0 and 100.0 (median [IQR]: 40.0 [21.0 – 72.5]).

**Table 1. Demographics and characteristics of the study participants at baseline.**

| Variable | HC included in ≥ 1 analysis (n = 35) | PwMS included in ≥ 1 analysis (n = 93) |
|---|---|---|
| Female, n (%) | 19 (54) | 57 (61) |
| Age, years, median (IQR); range | 54.0 (45.0 – 62.5); 28.0 – 75.0 | 56 (46 – 63); 26.0 – 77.0 |
| Height, m, median (IQR); range | 1.7 (1.6 – 1.8); 1.5 – 1.9 | 1.7 (1.6 – 1.8); 1.5 – 1.9 |
| Weight, kg, median (IQR); range | 78.0 (69.5 – 89.5); 60.0 – 130.0 | 75 (65 – 93); 46 – 142 |
| MS diagnosis, n (%) | | |
| Relapsing-remitting MS | NA | 65 (70) |
| Secondary progressive MS | NA | 15 (16) |
| Primary progressive MS | NA | 13 (14) |
| Time since MS symptom onset, years, median (IQR); range | NA | 18 (11 – 27); 3 – 53 |
| Time since MS diagnosis, years, median (IQR); range | NA | 14 (7 – 19); 1 – 39 |
| EDSS | | |
| Median (IQR); range | NA | 5.0 (4.0 –6.0); 1.5 – 6.5 |
| 0 - 3.5, n (%) | NA | 23 (25) |
| 4.0 - 5.5, n (%) | NA | 35 (38) |
| 6.0 - 6.5, n (%) | NA | 35 (38) |
| T25FW, s, median (IQR); range | NA | 5.4 (4.5 – 8.2); 2.9 – 18.6 |
| MSWS-12 version 2 transformed total score, median (IQR); range | NA | 40.0 (21.0 – 72.5); 0.0 – 100.0 |
| MSIS-29 | | |
| Physical subscale, median (IQR); range | NA | 27 (13 – 41.5); 0 – 73 |
| Psychological subscale, median (IQR); range | NA | 22 (25 – 11); 0 – 75 |

EDSS = Expanded Disability Status Scale; HC = healthy control; IQR = interquartile range; MSIS-29 = 29-item Multiple Sclerosis Impact Scale; MS = multiple sclerosis; MSWS-12 = Multiple Sclerosis Walking Scale; NA = not applicable; PwMS = people with multiple sclerosis; SD = standard deviation; T25FW = Timed 25-Foot Walk.

## Overall detection performance of gait events

### Initial contact

Structured in-lab overground and treadmill 2MWTs, performed during the second on-site visit, showed excellent median F1 scores across the six smartphone locations (HC: ≥ 99.0% in HC and ≥ 99.0% in PwMS; Figure 4a and Supplementary Table S1). On structured real-world walking, median (IQR) F1 scores consistently remained high, with HC at 100% (99.9% – 100%) and PwMS at 99.9% (99.0% – 100%). When comparing disease severity groups, the severe cohort showed the lowest F1 score of 99.3% (97.4% – 100%; Figure 4a). Unstructured real-world walking activities involving free walking yielded a F1 score of 97.8% (96.2% – 98.8%) in HC and 94.4% (90.4% – 96.6%) in PwMS, with a trend of higher F1 scores in less impaired patients (Figure 4a). F1 scores varied from 91.8% (86.6% – 94.7%) in severe PwMS to 96.0% (9.40% – 97.6%) in mild PwMS. Although there was a slight performance drop on unstructured real-world walking compared to other test conditions, the accuracy of the step detection remained very high.

Precision and recall scores were also high (Supplementary Table S1). Across test conditions and smartphone wear locations, median precision scores were ≥ 97.4% in HC and ≥ 93.0% in PwMS. Median recall scores were ≥ 98.7% in HC and ≥ 96.4% in PwMS.

Detailed statistics—including medians, means, quantiles, and interquartile ranges for the F1, precision, and recall scores—categorized by smartphone location, cohort, and test conditions are reported in Supplementary Table S1.

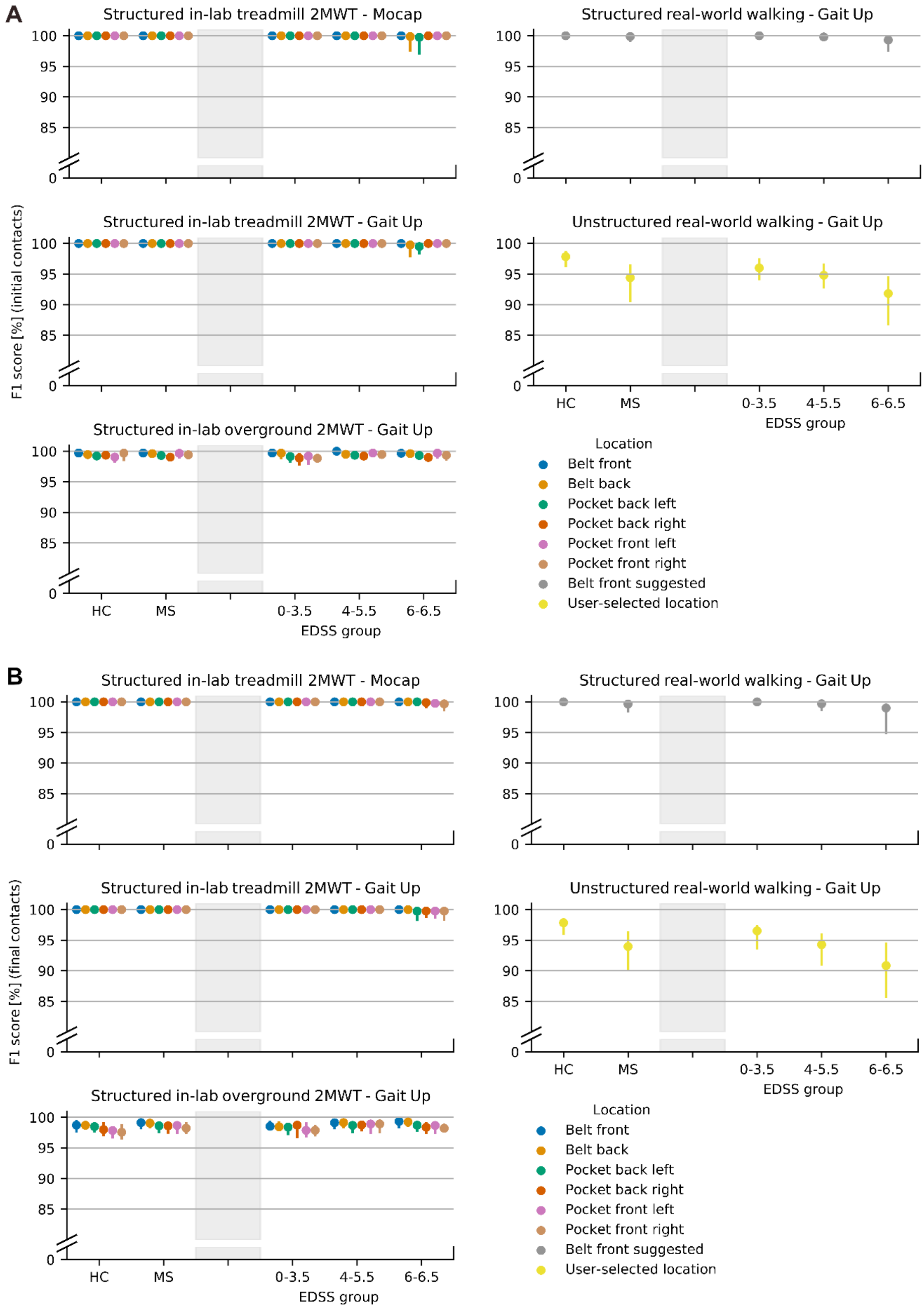

**Figure 4. Median F1 score (IQR) by smartphone wear location, cohort, and test setting.** (A) F1 scores are shown for initial contacts. (B) F1 scores are shown for final contacts. In both panels, note the break in the y-axes (F1 scores) between 5% and 85%, which was introduced for better readability. 2MWT = Two-Minute Walk Test; EDSS = Expanded Disability Status Scale; IQR = interquartile range; mocap = motion capture system.

## Final contact

Similar findings were obtained for the final contacts. Both structured in-lab overground and treadmill 2MWTs showed very high median F1 scores across smartphone wear locations (HC: ≥ 97.6%; PwMS: ≥ 98.2%; Figure 4b and Supplementary Table S2), indicating a 2% drop in median F1 scores on the structured in-lab overground 2MWT compared with initial contacts (final vs initial contact: ≥ 97.6% vs ≥ 99.0% for HC; ≥ 98.2% vs ≥ 99.0% for PwMS). Consistently high median (IQR) F1 scores were also observed on structured real-world walking (HC: 100% [99.8% – 100%]; PwMS: 99.6% [98.2% – 100%]) (Figure 4b). Across disease severity groups, F1 scores varied from 99.0% (94.7% – 99.7%) in severe PwMS to 100% (100%–100%) in mild PwMS. High F1 scores were also obtained on unstructured real-world walking (HC: 97.8% [95.8% – 98.6%]; PwMS: 94.0% [90.0% – 96.5%] PwMS; Figure 4b), with a trend of higher F1 in less impaired PwMS (severe PwMS: 90.9% [85.5% – 94.6%]; mild PwMS: 96.5% [93.4% – 97.5%]).

Precision and recall scores were equally high across test conditions and smartphone wear locations. Median precision scores were ≥ 97.5 for HC and ≥ 93.6% for PwMS. Median recall scores were ≥ 98.1% for HC and 95.4% for PwMS.

Detailed statistics—including medians, means, quantiles, and interquartile ranges for the F1 score, precision score, and recall score—categorized by smartphone location, cohort, and test conditions are reported in Supplementary Table S2.

## Temporal precision of gait event detection

### Initial contact

The temporal errors for initial contact were comparable between those recorded on the in-lab walking activities and those recorded on the real-world walking activities, as well as between HC and PwMS (Figure 5a and Supplementary Table S3). Both structured in-lab overground and treadmill 2MWTs indicated that the median errors across the six smartphone locations in HC and PwMS ranged between 0.02 s – 0.08 s (Figure 5a). Notably, smartphone locations on the front belt, back belt, and front pockets consistently yielded lower median errors, ranging from 0.02 s – 0.07 s, compared to those in back pockets, which exhibited median errors between 0.05 s – 0.08 s. Few smartphone wear locations, such as the front belt, back belt and front pockets, showed a mild trend of larger temporal errors being associated with higher EDSS scores; however, no such trend was observed for the back pockets (Figure 5a).

Temporal errors on structured and unstructured real-world walking activities were similar (Figure 5a). On structured real-world walking, median (IQR) errors were 0.05 s (0.05 s – 0.06 s) in HC and 0.06 s (0.05 s – 0.06 s) in PwMS, and on unstructured real-world walking activities they were 0.06 s (0.05 s – 0.06 s) and 0.07 s (0.06 s – 0.07 s), respectively. Of note, temporal errors remained stable with increasing EDSS scores on both structured and unstructured real-world walking (Figure 5a). Detailed statistics (mean and 95% CI) on temporal errors for initial contacts are reported in Supplementary Table S3.

### Final contact

Temporal errors observed for final contacts were generally in the same range as those observed for initial contacts (Figure 5b and Supplementary Table S4). However, final contact errors varied across smartphone locations on in-lab walking activities. In HC, but not PwMS, smartphones carried in the front belt, back belt, and front pockets consistently demonstrating lower temporal than smartphone placed in back pockets (median temporal error range: 0.02 s – 0.06 s vs 0.04 s – 0.08 s in HC; -0.01 s to 0.05 s vs 0.01 s – 0.05 s in PwMS) (Figure 5b).

In the real-world setting, temporal errors were similar for both structured and unstructured real-world walking activities, although the variability of these errors was greater on the unstructured walking activity. On structured real-world walking, median (IQR) temporal errors were 0.08 s (0.08 s – 0.09 s) in HC and 0.05 s (0.04 s – 0.07 s) in PwMS. On unstructured real-world walking activities, temporal errors were 0.07 s (0.06 s – 0.08 s) in HC and 0.06 s (0.05 s – 0.07 s) in PwMS (Figure 5b).

As the severity of the disease progressed, the temporal errors tended to improve in both in-lab and real-world settings. However, the variability in these errors also increased with higher EDSS scores, indicating that more severe cases may lead to less consistent results (Figure 5b). Detailed statistics (mean and 95% CI) on temporal errors for final contact are reported in Supplementary Table S4.

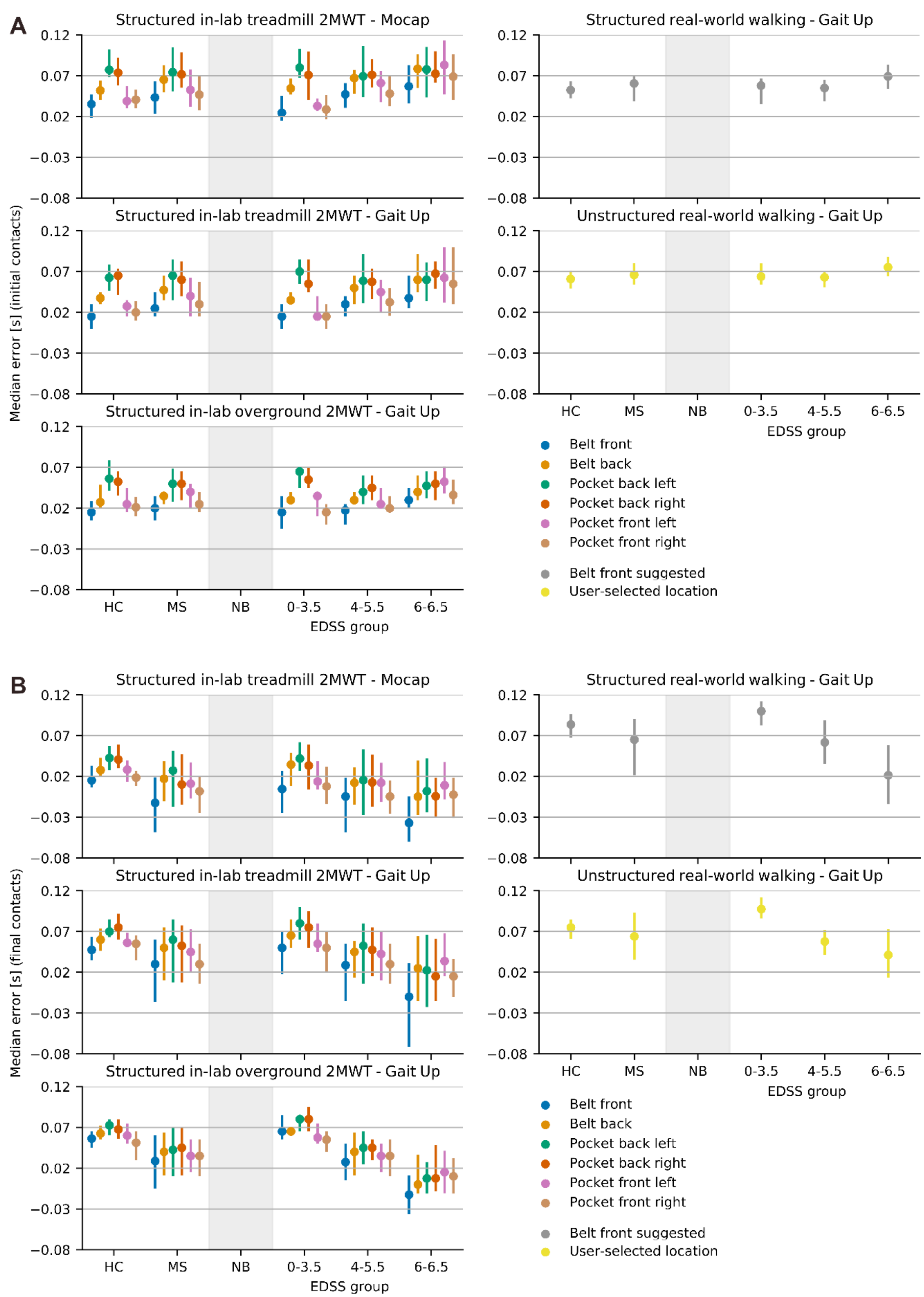

**Figure 5. Median temporal error (median [IQR]) by smartphone wear location, cohort, and test condition.** Median temporal errors for initial contacts (A) and final contacts (B) were aggregated across participants by computing the median and IQR across participants. 2MWT = Two-Minute Walk Test; EDSS = Expanded Disability Status Scale; IQR = interquartile range; mocap = motion capture system.

# Effect of potential confounding factors on gait event detection

## Initial contact

The impact of the examined confounding factors (i.e., age and gender, use of walking aid, indoors versus outdoors setting, and disease status and severity) on F1 score is minimal. None of the factors reach statistical significance, as evidenced by distributions crossing zero (Figure 6a). While statistically non-significant, the strongest differences were identified for disease status and severity, and the comparison between indoor and outdoor settings. In both scenarios, a decrease of about 1.7% – 2% in the median F1 score was observed for disease severity, as well as when participants walked indoors compared to their outdoor walking. More detailed statistics, including parameter distributions, are presented in Supplementary Table S5.

## Final contact

Results for the final contact are highly consistent with those reported for the initial contacts and are summarized in Figure 6b and Supplementary Table S6.

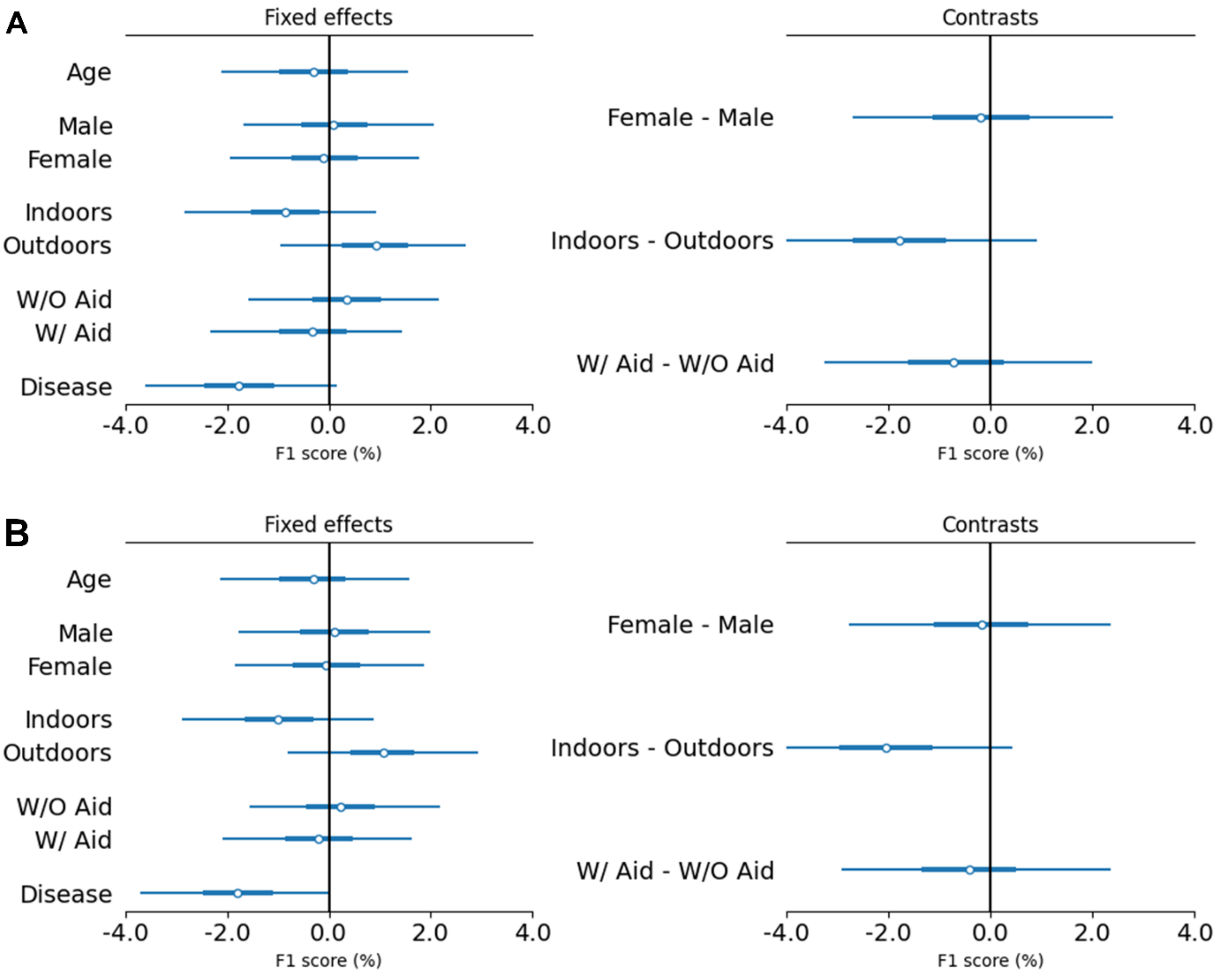

**Figure 6. Parameter estimates of Bayesian model examining the influence of different factors on the F1 score.** The forest plots depict the credible intervals of both the fixed effects and contrast of the Bayesian model for initial contacts (A) and final contacts (B). The central points are the estimated posterior median, the thick lines are the IQR range (0.25-0.75), and the thin lines represent the highest density intervals (94%). Contrast depicts the difference in the fixed effects between related factors such as sex (female vs male), test setting (indoors vs outdoors), are use of walking aids (with vs without walking aids). W/ Aid = with walking aids or orthotics; W/O Aid = without walking aids or orthotics.

# Discussion

This work introduces a gait pipeline with adaptive step detection that accurately identifies gait events across different walking activities in both HC and PwMS with varying disease severity. The results demonstrated the gait pipeline's ability to detect both initial and final contacts with high precision and high recall across multiple test conditions (controlled gait laboratory setting vs real-world scenarios), different smartphone sensor positions, and different real-world environments (indoor vs outdoor, but also with or without use of walking aids), expanding on previous research that focused mainly on validating the detection of initial contacts [29, 30]. These results confirm the robust technical validity of the pipeline and its potential for real-world unsupervised gait assessment. Importantly, this gait pipeline addresses one of the key challenges in advancing real-world gait analysis, the translation of wearable algorithms designed for and validated in controlled indoor settings to real-world unsupervised settings [33, 59].

## Overall detection performance of gait events

Our gait pipeline demonstrated high precision in identifying both initial and final contacts. Its performance in identifying initial contacts is comparable to that of the best-performing algorithms [37, 52, 60-63], as recently evaluated by Micó-Amigo et al. [64]. When tested on unstructured real-world walking activities in PwMS, it achieved a mean (95% CI) precision of 93% (88.9% – 95.9%), which is comparable to the 92% (91% – 93%) reported by Micó-Amigo et al. [64]. Additionally, our pipeline showed a higher recall than the best-performing algorithm assessed by Micó-Amigo et al. [64] (mean [95% CI] recall: 96.4% [93.4% – 98.4%] versus 82% [81% – 84%]). This greater overall performance compared with the work by Micó-Amigo et al. [64] was achieved even though the technical validation of our pipeline was conducted in more challenging conditions that could have potentially led to lower precision and recall metrics: (I) longer data collection periods (2 weeks vs 2.5 hours) and resulting larger number of gait bouts, (II) a larger cohort of more severely impaired PwMS who are more likely to use walking aids or orthotics (mean [SD] baseline EDSS: 4.75 [1.46] vs 3.5 [1.7]), (III) additional sensor positions (6 devices vs 1 device), and (IV) less controlled settings (both in-lab and real-world structured and unstructured walking activities vs specific tasks in real-world setting). Consequently, the data used for the technical validation of our gait pipeline captured a wider range of real-world walking variability, underscoring its robustness.

Previous research has highlighted the challenge of comparing results obtained in controlled settings with those from unsupervised, real-world settings due to differences in patient behaviors and movement complexity in natural surroundings [33, 65]. Remarkably, our gait pipeline consistently demonstrated high performance in both in-lab and real-world settings, regardless of the type of population (pathological and non-pathological) and disease stage. As can be expected, F1 scores were slightly lower on the unstructured real-world walking activities due to a greater number of shorter gait bouts lasting less than 10 s as well as a greater proportion of transitional phases of gait (e.g., initiation and termination of gait), turns, and changes in walking speed, all of which may negatively impact step detection performance. However, the reduction in F1 scores was minimal (approximately 2% points in HC and 6% points in PwMS), and the F1 scores remained higher than those reported for previously published algorithms [64].

The pipeline's improved performance can be explained by the novel step detection algorithm, which uses an adaptive wavelet to enhance the detection of initial and final contacts in different settings, and the robust anatomical reference system identification. The wavelet scale, sign, and the axis on which the wavelet is applied are determined based on the input acceleration signal. This adaptability by design allows the pipeline to increase its robustness in tackling several challenges, including significant variability in signal characteristics, intrinsic gait factors, such as population differences (e.g., healthy adults versus patients with various conditions and stages of disease), variations in walking speed (e.g., slow versus fast), differences in test conditions (e.g., supervised vs unsupervised, in-lab or on-site vs real-world or remote, structured vs unstructured walking activities), differences in contextual factor (e.g., indoors vs outdoors), and the use of walking aids. Together, these results suggest the potential utility of the gait pipeline for future assessments of pathological gait in natural and unobtrusive settings.

## Temporal precision of gait event detection

We found that temporal errors for initial contact were similar in magnitude to those for final contact. Significantly, the constant positive bias observed in both initial and final contacts is not expected to affect the computation of gait measures, such as stance and swing. This suggests the gait pipeline detects both types of gait events with a high degree of temporal accuracy.

For both initial and final contacts, temporal errors were ≤ 0.08 s and hence roughly 10x smaller than the tolerance window of 0.5 s. These errors remained relatively stable across various settings, including in-lab and real-world settings, structured and unstructured walking activities, different smartphone wear locations, and varying levels of disability. However, some small differences were noted in specific cases. For instance, smartphones placed in the back pocket tended to exhibit slightly larger temporal errors compared to other smartphone wear locations, likely due to increased inference from the soft tissue artifacts [66]. Furthermore, real-world walking activities tended to show larger temporal errors than in-lab walking activities.

Nonetheless, temporal errors for real-world walking (initial contact: 0.05 s – 0.07 s; final contact: 0.05 s – 0.08 s) fell within the same range as previously reported for CWT-based algorithms applied to in-lab gait data (≤ 0.10 s) [37, 64, 67]. However, previous assessments of temporal errors in identifying final contact have mainly been conducted in other patient populations, such as people with Parkinson’s disease [37] or hemiplegia [67]. However, as gait is differently impacted in PwMS compared to people with Parkinson’s disease or hemiplegia, a direct comparison of temporal errors across these two populations is difficult.

All these differences in temporal errors lie within the margin of uncertainty that should be considered when interpreting time differences in gait events. This uncertainty stems from the measurement error inherent in reference measurement systems and can be as large as 0.04 s. Specifically, the event detection algorithms incorporated in the Gait Offline Analysis Tool (GOAT) software package—employed to establish the reference initial and final contacts referenced in this paper—(see section “Supplementary appendix” for further details) reported temporal errors as large as 0.03 s in impaired population such PwMS or people with stroke [68, 69].

## Effect of potential confounding factors on gait event detection

While minor variations were observed, the overall detection performance remained robust and reliable across different factors, including disease severity (the pipeline showed good performance in detecting gait events in PwMS with either mild, moderate, or severe levels of MS-related disability) and real-world environments (the pipeline maintained high performance in both indoor and outdoor settings).

After examining the information from the gait diary on how the structured real-world walking was performed, a Bayesian analysis found that factors like indoor versus outdoor setting, use of walking aids, age, gender, population type, and disease stage had minimal and statistically non-significant impact on the F1 score. Age and gender did not affect the percentage difference in F1 score for initial and final contacts. The median F1 score was about 2% lower in indoor compared to outdoor settings for initial and final contacts. This small difference may be due to the more confined nature of indoor settings [34], potentially leading to a higher frequency of shorter gait bouts and an increased presence of transitional phases of gait and turns. Using walking aids can also cause changes in walking patterns, which in turn may directly impact the pipeline's performance [70, 71]. The performance of our gait pipeline was marginally reduced by 0.7% for initial contacts and 0.4% for final contacts in PwMS who used walking aids. Nonetheless, this reduction in performance did not reach statistical significance, indicating that while there is a notable impact, the use of walking aids does not substantially compromise the overall performance of our gait pipeline. Moreover, our pipeline maintains high performance across varying levels of MS-related disability or MS disease severity. A slight drop of approximately 2% in the median F1 score as disability increased; however, this change did not reach statistical significance, reaffirming the pipeline's robustness regardless of severity.

These findings indicate that the gait pipeline's performance remained consistent regardless of intrinsic gait factors, real-world environment and test conditions. Even with a slight drop in performance, our pipeline performed to a very high level of accuracy. The demographics and anthropometric characteristics of the enrolled PwMS exhibited considerable diversity. Specifically, PwMS with various clinical phenotypes of MS (i.e., relapsing-remitting MS, primary progressive MS, and secondary progressive MS), along with varying levels of disability (i.e., mild, moderate, and severe cohorts), were included in the analyses. This heterogeneity within the study population establishes a robust framework for objectively assessing the performance of the gait pipeline in MS. Additionally, it enhances the potential for generalizing the findings to a broader MS population. Furthermore, the pipeline has the added benefit of not requiring any knowledge of the exact sensor location and orientation of the smartphones, unlike many previously validated algorithms [36, 37, 52, 64, 65]. Consequently, our gait pipeline has less stringent requirements for future study participants or patients on how to carry the smartphones (or other IMU sensors) while performing the gait assessments.

## Limitations

This study comes with several limitations. First, the gait pipeline's performance in detecting gait events during turns could not be evaluated due to the absence of motion capture capabilities during turns—the motion capture system was only available for the structured in-lab treadmill 2MWT, which did not involve turns—and due to difficulties in consistently detecting gait events during turns with Gait Up IMUs. Second, there were challenges related to participant adherence, data loss, and synchronization problems. Nonetheless, the sample size available for the analyses was large enough (79 – 87 of the 100 PwMS enrolled and 32 – 35 of the 35 HC enrolled, depending on the analysis) to support the robustness of our gait pipeline. Third, the real-world walking activities were performed with a single smartphone device. Consequently, a within-subject comparison of different smartphone wear locations was not possible for these activities.

## Conclusions

This study demonstrates the technical validity and robustness of a gait pipeline that uses an adaptive step detection algorithm for use in PwMS, highlighting its potential for real-world gait assessment. Crucially, the pipeline performed with high accuracy across different test conditions and real-world environments, varying levels of disease severity, and different demographic characteristics. These findings expand on previous research by not only validating the detection of initial contacts but also the detection of final contacts. Such accurate step detection is essential for reliably estimating digital gait measures, such as cadence, step velocity, stance, step symmetry, and gait variability. It is, therefore, expected that the findings will advance our understanding of gait analysis and contribute to this field's existing knowledge. By developing a new and robust gait pipeline that offers flexibility to the setting in which gait data are captured and where the smartphone is worn, could pave the way for an assessment of pathological gait that is not only accurate and reliable but also easily deployable in the real world.

# Declarations

## Ethics approval and consent to participate

All participants provided written informed consent to participate in the study, which was approved by the UK Health Research Authority (IRAS Project ID: 302099).

## Consent for publication

Not applicable

## Data availability

Requesting the data underlying this publication requires a detailed, hypothesis-driven statistical analysis plan collaboratively developed by the requester and company subject matter experts. Such requests should be directed to dbm.datarequest@roche.com for consideration. Anonymized records for individual patients across >1 data source external to Roche cannot, and should not, be linked due to a potential increase in the risk of patient reidentification.

## Competing interests

The author(s) declared the following potential conflicts of interest concerning the research, authorship, and/or publication of this article: LA is an employee of and shareholder in F. Hoffmann-La Roche Ltd. DS is a consultant for F. Hoffmann-La Roche Ltd via Capgemini Engineering. MP is a contractor for Roche Polska Sp. z o.o. RK is a contractor for F. Hoffmann-La Roche Ltd. NN is a contractor for Roche Polska Sp. z o.o. GGC is an employee of the University of Plymouth, which received research funding from F. Hoffman-La Roche Ltd. LB is an employee of the University of Plymouth, which received research funding for this study from F. Hoffmann-La Roche Ltd. PSG is an employee of the University of Plymouth. RH has received indirect funding from F. Hoffmann-La Roche Ltd via a University of Plymouth Enterprise Ltd contract. JF is an employee of the University of Plymouth, which received research funding for this study from F. Hoffmann-La Roche Ltd. JM is an employee of the University of Plymouth, which received research funding for this study from F. Hoffmann-La Roche Ltd. JH or affiliated institutions have received either consulting fees, honoraria, support to attend meetings, clinical service support, or research support from Acorda, Bayer Schering Pharma, Biogen Idec., Brickell Biotech, F. Hoffmann-La Roche Ltd, Global Blood Therapeutics, Sanofi Genzyme, Merck Serono, Novartis, Oxford Health Policy Forum, Teva, and Vantia. LC is an employee of and shareholder in F. Hoffmann-La Roche Ltd. MDR is a contractor for F. Hoffmann-La Roche Ltd. MZ is an employee of and shareholder in F. Hoffmann-La Roche Ltd.

## Funding

This research was funded by F. Hoffmann-La Roche Ltd, Basel, Switzerland.

## Authors’ contributions

LA: Conceptualization, Data Curation, Formal analysis, Methods, Software, Visualization, Writing – Original Draft, Review & Editing

DS: Conceptualization, Data Curation, Formal analysis, Methods, Software, Visualization, Writing - Review & Editing

MP: Data Curation, Formal analysis, Methods, Software, Visualization, Writing - Review & Editing

RK: Data Curation, Formal analysis, Methods, Software, Visualization, Writing - Review & Editing

NN: Data Curation, Formal analysis, Software, Visualization, Writing - Review & Editing

GGC: Data Curation, Investigation, Writing - Review & Editing

LB: Data Curation, Investigation, Writing - Review & Editing

PSG: Data Curation, Investigation, Writing - Review & Editing

RH: Data Curation, Investigation, Writing - Review & Editing

JF: Conceptualization, Data Curation, Investigation, Writing - Review & Editing

JM: Conceptualization, Data Curation, Investigation, Supervision, Writing - Review & Editing

JH: Conceptualization, Data Curation, Investigation, Writing - Review & Editing

LC: Conceptualization, Funding acquisition, Writing - Review & Editing

MDR: Conceptualization, Formal analysis, Methods, Project administration, Supervision, Writing - Review & Editing

MZ: Conceptualization, Formal analysis, Methods, Project administration, Supervision, Writing - Review & Editing

## Acknowledgements

The authors would like to thank all study participants and their families. The authors would also like to thank the following current and former colleagues at F. Hoffmann-La Roche Ltd for their contributions to and support of the study: Pawel Biernacki, Matthias Bobst, Marco Bosatta, Sandro Fritz, Francisco Gavidia, Petra Hauser, Wiktoria Kasprzyk, Dominik Kedziora, Piotr Konorski, Kostas Kritsas, Hugo Le Gall, Dominik Lenart, Michael Lindemann, Florian Lipsmeier, David Meyer, Madalina Ogica, Emanuele Passerini, Szymon Piwowar, Elena Spyridou, and Lukasz Syzmankiewicz. The authors would also like to thank colleagues at the University of Plymouth for their contributions and support to the study: Michael Paisey, Tanya King, Joanne Lind, Carol Lunn, and Maureen Pedder. Editorial and writing support was provided by Sven Holm, PhD, contractor for F. Hoffmann-La Roche Ltd.

*Supplementary appendix to*

# Adaptive and robust smartphone-based step detection in multiple sclerosis

Lorenza Angelini,[1] Dimitar Stanev,[1] Marta Płonka,[2] Rafał Klimas,[1] Natan Napiórkowski,[2] Gabriela González Chan,[3] Lisa Bunn,[3] Paul S Glazier,[3] Richard Hosking,[4] Jenny Freeman,[3] Jeremy Hobart,[5] Jonathan Marsden,[3] Licinio Craveiro,[1] Mike D Rinderknecht,[1*] and Mattia Zanon[1*]

* Shared last author

[1]F. Hoffmann-La Roche Ltd, Basel, Switzerland

[2]Roche Polska Sp. z o.o., Warsaw, Poland

[3]School of Health Professions, University of Plymouth, Plymouth, UK

[4]University of Plymouth Enterprise Ltd, Plymouth, UK

[5]Peninsula Medical School, Faculty of Health, University of Plymouth, Plymouth, UK

**Corresponding author**

Mike D Rinderknecht

mike.rinderknecht@roche.com

## Data synchronisation

The work by Stanev et al. [1] presents a methodology for synchronising data across various measurement systems, enhancing the accuracy and reliability of data acquisition in clinical settings. Smartphones and Gait Up IMUs were synchronised using a mechanical perturbation method, where devices were subjected to a rocking motion to generate identifiable gyroscope signal patterns, allowing for temporal offset correction. A vibration-based method was also employed to synchronise smartphones and Gait Up IMUs with mocap system, using a vibrating motor to generate pulse sequences for alignment. Despite potential lags introduced by the vibrating motor's inertia or pulse sequence delays, post hoc synchronisation techniques, such as the force-based lag estimation method, were utilised to correct these discrepancies. This method leverages the relationship between ground reaction forces (GRFs) and center of mass acceleration during walking or standing, using accelerometer readings from a smartphone to estimate and rectify lags, thereby improving data quality.

## Gait event detection from the reference tools

The mocap and GRF data were processed using the Gait Offline Analysis Tool (GOAT) software package, which was developed by manufacturer [2]. This software offers built-in algorithms that can accurately identify initial and final contacts. GOAT features two distinct event detection algorithms, namely the kinematic-based and GRF-based algorithms, making it a reliable tool for gait analysis in healthy individuals and those with pathological conditions [3]. The kinematic-based algorithm, also known as the coordinate-based algorithm, detects gait events based on the positions of sacrum and foot markers filtered with a 2nd order Butterworth filter and cut-off frequency of 6Hz. This algorithm was validated against the events identified from the vertical GRF, which is considered the gold standard for such analysis. Comprehensive details of the algorithmic implementation have been previously described [4]. The GRF-based algorithm, on the other hand, relies on force data that have been converted to center-of-pressure (COP) data. This conversion enables the identification of gait events using the characteristic COP “butterfly” pattern that emerges over time. The initial and final contacts obtained from the COP profile were validated against those based on kinematic data, which were assumed to represent “true” gait events [5]. By default, the gait events used in the subsequent analyses were identified from the gold standard using the GRF-based algorithm. However, when a walking aid was used, the kinematic-based method was employed to identify gait events.

The gait events were obtained from the silver standard using a developed gait analysis package, which has been previously validated against motion capture systems in both young and elderly individuals [6-9] and in limping walking conditions across a range of walking speeds between 0.9 and 2.0 m/second [10]. Foot-worn IMUs were primarily used to detect gait events (e.g., initial and final contacts). In contrast, the IMU attached to the lower back was used for more accurate turn detection.

## Verification of sagittal and frontal planes through autocorrelation

The frontal and sagittal planes were initially identified through principal component analysis (PCA). However, at times PCA may misidentify these two planes. The correct identification of the frontal and sagittal planes can be confirmed by verifying their directionality, i.e., the antero-posterior direction for the frontal plane and the medio-lateral direction for the sagittal plane.

Verification of the antero-posterior and medio-lateral direction was achieved with an autocorrelation-based algorithm. First, the autocorrelations of the smartphone sensor signals along the three directions (i.e., vertical, antero-posterior, and medio-lateral direction) were computed ("initial autocorrelations"). To reduce noise and remove spurious peaks from these "initial autocorrelations", additional processing steps were implemented. These included calculating the "sum autocorrelation" as the sum of the three "initial autocorrelations", computing the "norm autocorrelation" as the square root of the sum of the squares of the three "initial autocorrelations", and finally, combining the "sum autocorrelation" and the "norm autocorrelation" to generate a "combined autocorrelation" signal that is less noisy than the "initial autocorrelations".

The location of the first peak in this "combined autocorrelation" signal provides a robust estimate of the step duration. This location was used for subsequent verification of antero-posterior and medio-lateral directions. Specifically, this verification step evaluated peaks at this same location but in the initial autocorrelation signals instead. Given the physical properties of walking, the anterior-posterior acceleration is biphasic, and hence, its initial autocorrelation signal should show a first positive peak. The mediolateral acceleration of walking, by comparison, is monophasic. Consequently, the first peak in its initial autocorrelation signal should be negative (note: the vertical direction was already correctly identified during the gravity alignment step). Thus, the anterior-posterior and medio-lateral directions were correctly identified by the PCA if the first peak in the initial autocorrelation signal of the antero-posterior direction is positive and the corresponding peak in the initial autocorrelation signal of the medio-lateral direction is negative. If both first peaks are negative, the antero-posterior direction can be correctly identified by the initial autocorrelation signal that has the first peak with the highest absolute value (as it is the direction of motion). If, however, the antero-posterior peak is negative and the medio-lateral peak is positive, it suggests that the PCA misidentified the antero-posterior and medio-lateral directions (i.e., the frontal and sagittal planes) were misidentified, and the directions (i.e. planes) were swapped accordingly.

## Bayesian model

The model's outcome variable was the F1 Score, which ranged from 0 to 100 or from 0 to 1 when normalised. Given the bounded nature of the outcome variable, a beta distribution between 0 and 1 was employed. The beta distribution, which is utilised for modelling the F1 score, is parameterised by the α and β parameters, which are positive. The α and β parameters are parametrised in terms of μ and k as follows (also known as beta regression reparametrisation):

$$\alpha = \mu \cdot k$$

$$\beta = (1 - \mu) \cdot k \text{ with } 0 < \mu < 1$$

A mathematical description of the Bayesian beta regression model is presented in Supplementary Figure S2.

The parameter μ was utilised to link the predictor variables to the distribution parameters. Given the bounded constraint of μ (0 < μ < 1), a link sigmoid function (logit) was introduced, leading to the formulation of a generalised linear model. Parameters a and b represent the intercept and age parameters; s, and d denote the sex and disease parameters corresponding to their index variables; u accommodates subject intercepts to account for repeated measures within subjects; e and h represent the effects under examination (e.g., indoor/outdoor and walking aid). Furthermore, parameter d was constrained to behave as an ordinal predictor variable as a function of disease severity. Tables S1 and S2 summarise the contrast parameter estimates of the Bayesian model for the initial and final contacts, respectively.

# Supplementary figures and tables

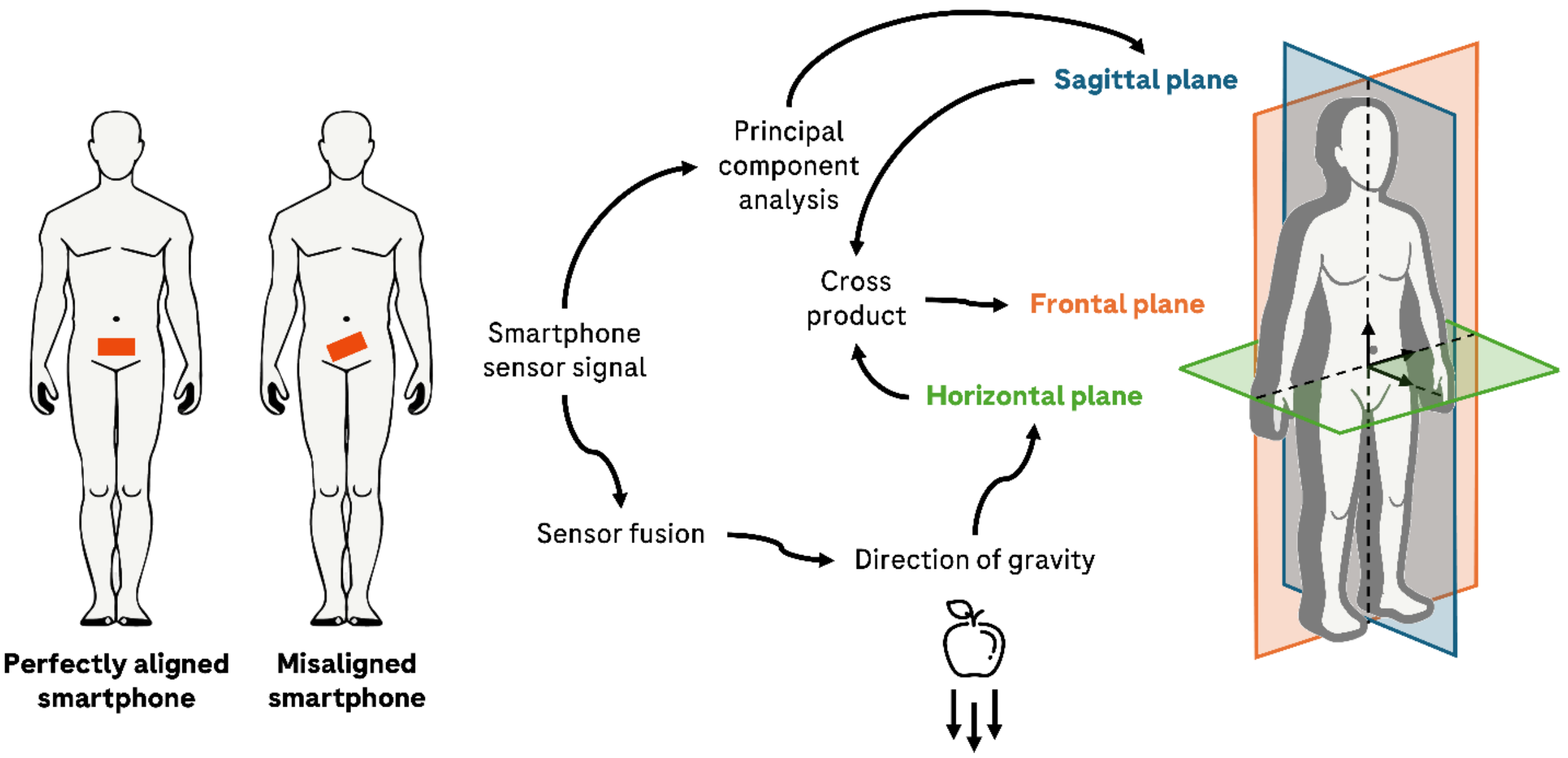

**Supplementary Figure S1. Alignment with gravity and anatomical reference frame.** Smartphone accelerometers that are not aligned with the direction of movement may negatively impact the performance of the step detection algorithm. Thus, aligning the smartphone accelerometer and gyroscope signal to the direction of gravity and the anatomical reference frame that is agnostic to the smartphone's orientation is required to ensure the algorithm's ability to handle differently oriented smartphone devices. The direction of gravity, or horizontal plane, was identified through sensor fusion (i.e., Madgwick filter [11] applied to the smartphone accelerometer and gyroscope signals). The two remaining planes that together with the horizontal plane make up the orientation-agnostic anatomical reference frame were identified as follows: The sagittal plane (i.e., the direction of movement) was identified by conducting principal component analysis on the gravity-aligned accelerometer data and the frontal plane through computing the cross product of the horizontal and sagittal planes.

$$
\begin{gathered}
\mathrm{F1}_{obs}[i] \sim \mathrm{Beta}(\alpha[i], \beta[i]),\ i \in \{0, \ldots, N-1\} \\
\alpha[i] = \mu[i] \cdot \kappa \\
\beta[i] = (1 - \mu[i]) \cdot \kappa \\
\mathrm{logit}(\mu[i]) = a + b \cdot \bar{age} + s[sex_{idx}[i]] + d \cdot \sum_{j=0}^{disease_{idx}[i]-1} \delta_j + u[subject_{idx}[i]] + e[environment_{idx}[i]] + h[aid_{idx}[i]] \\
\kappa \sim \mathrm{HalfCauchy}(20) \\
a \sim \mathrm{Normal}(1, 1) \\
s = 1/100 \\
b \sim \mathrm{Normal}(0, s) \\
\mu_{sub} \sim \mathrm{Normal}(0, s) \\
\sigma_{sub} \sim \mathrm{HalfCauchy}(s) \\
s[j] \sim \mathrm{Normal}(0, s),\ j \in \{\mathrm{M}, \mathrm{F}\} \\
d \sim \mathrm{Normal}(0, s) \\
\delta[j] \sim \mathrm{Diriclet}(1),\ j \in \{\text{disease categories}\} \\
u[j] \sim \mathrm{Normal}(\mu_{sub}, \sigma_{sub}),\ j \in \{\text{subjects}\} \\
e[j] \sim \mathrm{Normal}(0, s),\ j \in \{\text{Indoor}, \text{Outdoor}\} \\
h[j] \sim \mathrm{Normal}(0, s),\ j \in \{\text{W/ Aid}, \text{W/O Aid}\}
\end{gathered}
$$

**Supplementary Figure S2. Complete description of the Bayesian model.**

**Supplementary Table S1. Algorithm performance for initial contact.**

| Cohort[a] | Smartphone location | Precision | | | | | | | | Recall | | | | | | | | F1 score | | | | | | | |
|---|---|---|---|---|---|---|---|---|---|---|---|---|---|---|---|---|---|---|---|---|---|---|---|---|---|
| | | Mdn | P05 | Q1 | Q3 | P95 | ws-IQR | Mean | 95% CI | Mdn | P05 | Q1 | Q3 | P95 | ws-IQR | Mean | 95% CI | Mdn | P05 | Q1 | Q3 | P95 | ws-IQR | Mean | 95% CI |
| ***Structured in-lab treadmill 2MWT – smartphone vs mocap*** | | | | | | | | | | | | | | | | | | | | | | | | | |
| HC | BB | 100 | 100 | 100 | 100 | 100 | 0 | 100 | 99.9 - 100 | 100 | 100 | 100 | 100 | 100 | 0 | 100 | 100 - 100 | 100 | 100 | 100 | 100 | 100 | 0 | 100 | 99.9 - 100 |
| | BF | 100 | 99.7 | 100 | 100 | 100 | 0 | 99.9 | 99.8 - 100 | 100 | 99.9 | 100 | 100 | 100 | 0 | 100 | 100 - 100 | 100 | 99.7 | 100 | 100 | 100 | 0 | 100 | 99.9 - 100 |
| | PB – left | 100 | 99.5 | 100 | 100 | 100 | 0 | 99.9 | 99.8 - 100 | 100 | 99.5 | 100 | 100 | 100 | 0 | 99.8 | 99.6 - 100 | 100 | 99.4 | 100 | 100 | 100 | 0 | 99.9 | 99.7 - 100 |
| | PB – right | 100 | 98.8 | 100 | 100 | 100 | 0 | 99.9 | 99.7 - 100 | 100 | 99.5 | 100 | 100 | 100 | 0 | 99.9 | 99.8 - 100 | 100 | 99.3 | 100 | 100 | 100 | 0 | 99.9 | 99.8 - 100 |
| | PF – left | 100 | 99.6 | 100 | 100 | 100 | 0 | 99.9 | 99.8 - 100 | 100 | 100 | 100 | 100 | 100 | 0 | 100 | 100 - 100 | 100 | 99.7 | 100 | 100 | 100 | 0 | 100 | 99.9 - 100 |
| | PF – right | 100 | 100 | 100 | 100 | 100 | 0 | 100 | 99.9 - 100 | 100 | 99.8 | 100 | 100 | 100 | 0 | 100 | 99.9 - 100 | 100 | 99.7 | 100 | 100 | 100 | 0 | 100 | 99.9 - 100 |
| Mild PwMS | BB | 100 | 99.2 | 100 | 100 | 100 | 0 | 99.9 | 99.6 - 100 | 100 | 99.1 | 100 | 100 | 100 | 0 | 99.9 | 99.7 - 100 | 100 | 99.3 | 100 | 100 | 100 | 0 | 99.9 | 99.7 - 100 |
| | BF | 100 | 99.6 | 100 | 100 | 100 | 0 | 99.8 | 99.6 - 100 | 100 | 99.5 | 100 | 100 | 100 | 0 | 99.9 | 99.7 - 100 | 100 | 99.7 | 100 | 100 | 100 | 0 | 99.9 | 99.6 - 100 |
| | PB – left | 100 | 99.2 | 100 | 100 | 100 | 0 | 99.8 | 99.5 - 100 | 100 | 98.1 | 100 | 100 | 100 | 0 | 99.8 | 99.5 - 100 | 100 | 99 | 100 | 100 | 100 | 0 | 99.8 | 99.5 - 100 |
| | PB – right | 100 | 99.6 | 100 | 100 | 100 | 0 | 99.9 | 99.6 - 100 | 100 | 98.8 | 100 | 100 | 100 | 0 | 99.8 | 99.6 - 100 | 100 | 99.3 | 100 | 100 | 100 | 0 | 99.8 | 99.6 - 100 |
| | PF – left | 100 | 100 | 100 | 100 | 100 | 0 | 100 | 99.9 - 100 | 100 | 99.6 | 100 | 100 | 100 | 0 | 99.9 | 99.8 - 100 | 100 | 99.6 | 100 | 100 | 100 | 0 | 99.9 | 99.9 - 100 |
| | PF – right | 100 | 99.6 | 100 | 100 | 100 | 0 | 100 | 99.9 - 100 | 100 | 100 | 100 | 100 | 100 | 0 | 100 | 100 - 100 | 100 | 99.8 | 100 | 100 | 100 | 0 | 100 | 100 - 100 |
| Moderate PwMS | BB | 100 | 99.6 | 100 | 100 | 100 | 0 | 99.9 | 99.7 - 100 | 100 | 99.3 | 100 | 100 | 100 | 0 | 99.8 | 99.6 - 100 | 100 | 99.2 | 100 | 100 | 100 | 0 | 99.9 | 99.7 - 100 |
| | BF | 100 | 98.3 | 100 | 100 | 100 | 0 | 99.6 | 99.2 - 100 | 100 | 98.1 | 100 | 100 | 100 | 0 | 99.6 | 99.2 - 100 | 100 | 98.4 | 100 | 100 | 100 | 0 | 99.6 | 99.2 - 100 |
| | PB – left | 100 | 96 | 100 | 100 | 100 | 0 | 99.3 | 98.5 - 100 | 100 | 90.1 | 99.8 | 100 | 100 | 0.2 | 98 | 95.9 - 100 | 100 | 91.6 | 99.8 | 100 | 100 | 0.2 | 98.5 | 97.1 - 100 |
| | PB – right | 100 | 96.8 | 100 | 100 | 100 | 0 | 99.5 | 98.9 - 100 | 100 | 93.9 | 100 | 100 | 100 | 0 | 99.2 | 98.6 - 99.9 | 100 | 95 | 99.9 | 100 | 100 | 0.1 | 99.4 | 98.8 - 99.9 |
| | PF – left | 100 | 97.9 | 100 | 100 | 100 | 0 | 98.6 | 96.3 - 100 | 100 | 97.7 | 100 | 100 | 100 | 0 | 98.6 | 96.2 - 100 | 100 | 98.2 | 100 | 100 | 100 | 0 | 98.6 | 96.2 - 100 |
| | PF – right | 100 | 99 | 100 | 100 | 100 | 0 | 99.8 | 99.6 - 100 | 100 | 98.8 | 100 | 100 | 100 | 0 | 99.8 | 99.7 - 100 | 100 | 99 | 100 | 100 | 100 | 0 | 99.8 | 99.7 - 100 |
| Severe PwMS | BB | 100 | 92.7 | 97.4 | 100 | 100 | 2.6 | 94.8 | 87.3 - 100 | 100 | 85.8 | 97.4 | 100 | 100 | 2.6 | 94.3 | 86.7 - 100 | 99.9 | 89.1 | 97.4 | 100 | 100 | 2.6 | 94.6 | 87 - 100 |
| | BF | 100 | 97.5 | 99.4 | 100 | 100 | 0.6 | 97.8 | 94 - 100 | 100 | 97.8 | 99.4 | 100 | 100 | 0.6 | 99.6 | 99.3 - 99.9 | 100 | 97.5 | 99.6 | 100 | 100 | 0.4 | 98.4 | 96 - 100 |
| | PB – left | 100 | 88 | 97.9 | 100 | 100 | 2.1 | 96 | 91.7 - 100 | 100 | 86.6 | 95.7 | 100 | 100 | 4.3 | 95.5 | 91.2 - 99.8 | 99.8 | 87.6 | 96.9 | 100 | 100 | 3.1 | 95.8 | 91.4 - 100 |
| | PB – right | 100 | 79.9 | 99 | 100 | 100 | 1 | 96.7 | 93.1 - 100 | 100 | 78.1 | 97.9 | 100 | 100 | 2.1 | 95.5 | 91.7 - 99.4 | 100 | 79 | 98 | 100 | 100 | 2 | 96.1 | 92.4 - 99.8 |
| | PF – left | 100 | 66.4 | 99.1 | 100 | 100 | 0.9 | 95.7 | 90.4 - 100 | 100 | 66.4 | 98.5 | 100 | 100 | 1.5 | 95.5 | 90.1 - 100 | 100 | 66.4 | 99.1 | 100 | 100 | 0.9 | 95.6 | 90.3 - 100 |
| | PF – right | 100 | 70.6 | 99.1 | 100 | 100 | 0.9 | 96.3 | 92.4 - 100 | 100 | 69.8 | 97.7 | 100 | 100 | 2.3 | 95.8 | 91.6 - 100 | 100 | 70.2 | 98.8 | 100 | 100 | 1.2 | 96 | 92 - 100 |
| All PwMS | BB | 100 | 95.8 | 100 | 100 | 100 | 0 | 98.4 | 96.2 - 100 | 100 | 95.8 | 100 | 100 | 100 | 0 | 98.2 | 96 - 100 | 100 | 95.8 | 100 | 100 | 100 | 0 | 98.3 | 96.1 - 100 |
| | BF | 100 | 97.5 | 100 | 100 | 100 | 0 | 99.1 | 98 - 100 | 100 | 97.8 | 100 | 100 | 100 | 0 | 99.7 | 99.5 - 99.9 | 100 | 97.5 | 99.8 | 100 | 100 | 0.2 | 99.3 | 98.6 - 100 |
| | PB – left | 100 | 90.9 | 100 | 100 | 100 | 0 | 98.5 | 97.1 - 99.8 | 100 | 87.6 | 99.5 | 100 | 100 | 0.5 | 97.8 | 96.2 - 99.3 | 100 | 88.6 | 99.6 | 100 | 100 | 0.4 | 98.1 | 96.6 - 99.5 |
| | PB – right | 100 | 94.9 | 100 | 100 | 100 | 0 | 98.8 | 97.7 - 99.9 | 100 | 91.6 | 99.7 | 100 | 100 | 0.3 | 98.3 | 97.1 - 99.5 | 100 | 93 | 99.8 | 100 | 100 | 0.2 | 98.5 | 97.4 - 99.7 |
| | PF – left | 100 | 97.8 | 100 | 100 | 100 | 0 | 98.1 | 96.3 - 100 | 100 | 97.3 | 100 | 100 | 100 | 0 | 98.1 | 96.2 - 100 | 100 | 97.9 | 99.8 | 100 | 100 | 0.2 | 98.1 | 96.2 - 100 |
| | PF – right | 100 | 97.2 | 100 | 100 | 100 | 0 | 98.8 | 97.7 - 100 | 100 | 96.2 | 100 | 100 | 100 | 0 | 98.7 | 97.4 - 100 | 100 | 97.1 | 99.9 | 100 | 100 | 0.1 | 98.8 | 97.5 - 100 |
| ***Structured in-lab treadmill 2MWT – smartphone vs Gait Up*** | | | | | | | | | | | | | | | | | | | | | | | | | |
| HC | BB | 100 | 100 | 100 | 100 | 100 | 0 | 100 | 100 - 100 | 100 | 100 | 100 | 100 | 100 | 0 | 100 | 100 - 100 | 100 | 100 | 100 | 100 | 100 | 0 | 100 | 100 - 100 |
| | BF | 100 | 100 | 100 | 100 | 100 | 0 | 100 | 99.9 - 100 | 100 | 100 | 100 | 100 | 100 | 0 | 100 | 100 - 100 | 100 | 99.9 | 100 | 100 | 100 | 0 | 100 | 100 - 100 |
| | PB – left | 100 | 100 | 100 | 100 | 100 | 0 | 100 | 99.9 - 100 | 100 | 99.5 | 100 | 100 | 100 | 0 | 99.8 | 99.6 - 100 | 100 | 99.5 | 100 | 100 | 100 | 0 | 99.9 | 99.7 - 100 |
| | PB – right | 100 | 100 | 100 | 100 | 100 | 0 | 99.9 | 99.8 - 100 | 100 | 99.7 | 100 | 100 | 100 | 0 | 100 | 99.9 - 100 | 100 | 99.9 | 100 | 100 | 100 | 0 | 100 | 99.9 - 100 |
| | PF – left | 100 | 99.6 | 100 | 100 | 100 | 0 | 100 | 99.9 - 100 | 100 | 100 | 100 | 100 | 100 | 0 | 100 | 100 - 100 | 100 | 99.7 | 100 | 100 | 100 | 0 | 100 | 99.9 - 100 |
| | PF – right | 100 | 100 | 100 | 100 | 100 | 0 | 100 | 100 - 100 | 100 | 100 | 100 | 100 | 100 | 0 | 100 | 100 - 100 | 100 | 100 | 100 | 100 | 100 | 0 | 100 | 100 - 100 |
| Mild PwMS | BB | 100 | 99.1 | 100 | 100 | 100 | 0 | 99.8 | 99.5 - 100 | 100 | 99 | 100 | 100 | 100 | 0 | 99.8 | 99.6 - 100 | 100 | 99.5 | 100 | 100 | 100 | 0 | 99.8 | 99.5 - 100 |
| | BF | 100 | 99.6 | 100 | 100 | 100 | 0 | 99.8 | 99.5 - 100 | 100 | 99.4 | 100 | 100 | 100 | 0 | 99.9 | 99.6 - 100 | 100 | 99.7 | 100 | 100 | 100 | 0 | 99.8 | 99.6 - 100 |
| | PB – left | 100 | 99.6 | 100 | 100 | 100 | 0 | 99.8 | 99.4 - 100 | 100 | 98 | 100 | 100 | 100 | 0 | 99.8 | 99.5 - 100 | 100 | 99 | 100 | 100 | 100 | 0 | 99.8 | 99.4 - 100 |
| | PB – right | 100 | 99.6 | 100 | 100 | 100 | 0 | 99.8 | 99.4 - 100 | 100 | 98.7 | 100 | 100 | 100 | 0 | 99.7 | 99.4 - 100 | 100 | 99.3 | 100 | 100 | 100 | 0 | 99.7 | 99.4 - 100 |
| | PF – left | 100 | 99.6 | 100 | 100 | 100 | 0 | 99.9 | 99.9 - 100 | 100 | 99.7 | 100 | 100 | 100 | 0 | 100 | 99.9 - 100 | 100 | 99.6 | 100 | 100 | 100 | 0 | 99.9 | 99.9 - 100 |
| | PF – right | 100 | 99.6 | 100 | 100 | 100 | 0 | 99.9 | 99.8 - 100 | 100 | 100 | 100 | 100 | 100 | 0 | 100 | 100 - 100 | 100 | 99.8 | 100 | 100 | 100 | 0 | 100 | 99.9 - 100 |
| Moderate PwMS | BB | 100 | 98.2 | 100 | 100 | 100 | 0 | 99.7 | 99.5 - 99.9 | 100 | 99.5 | 100 | 100 | 100 | 0 | 99.9 | 99.7 - 100 | 100 | 98.9 | 100 | 100 | 100 | 0 | 99.8 | 99.6 - 99.9 |
| | BF | 100 | 98.2 | 99.6 | 100 | 100 | 0.4 | 99.6 | 99.4 - 99.9 | 100 | 99 | 100 | 100 | 100 | 0 | 99.8 | 99.7 - 100 | 100 | 98.5 | 99.8 | 100 | 100 | 0.2 | 99.7 | 99.6 - 99.9 |
| | PB – left | 100 | 96.5 | 99.3 | 100 | 100 | 0.7 | 99.2 | 98.7 - 99.8 | 100 | 92.6 | 100 | 100 | 100 | 0 | 98.3 | 96.1 - 100 | 100 | 94.3 | 99.4 | 100 | 100 | 0.6 | 98.6 | 97.3 - 100 |
| | PB – right | 100 | 97.9 | 99.3 | 100 | 100 | 0.7 | 99.5 | 99.2 - 99.8 | 100 | 96.2 | 99.5 | 100 | 100 | 0.5 | 99.4 | 98.9 - 99.8 | 100 | 97.6 | 99.1 | 100 | 100 | 0.9 | 99.4 | 99.1 - 99.8 |
| | PF – left | 100 | 97.9 | 99.7 | 100 | 100 | 0.3 | 98.4 | 95.9 - 100 | 100 | 98.5 | 100 | 100 | 100 | 0 | 98.6 | 96.1 - 100 | 100 | 98.7 | 99.5 | 100 | 100 | 0.5 | 98.5 | 96 - 100 |
| | PF – right | 100 | 97.8 | 100 | 100 | 100 | 0 | 99.5 | 99.2 - 99.8 | 100 | 98.4 | 100 | 100 | 100 | 0 | 99.7 | 99.6 - 99.9 | 100 | 98.3 | 99.4 | 100 | 100 | 0.6 | 99.6 | 99.4 - 99.9 |
| Severe PwMS | BB | 100 | 93.2 | 97.8 | 100 | 100 | 2.2 | 94.6 | 86.5 - 100 | 100 | 86.2 | 98.4 | 100 | 100 | 1.6 | 94.2 | 86.1 - 100 | 99.8 | 89.6 | 97.7 | 100 | 100 | 2.3 | 94.4 | 86.3 - 100 |
| | BF | 100 | 92.4 | 99.5 | 100 | 100 | 0.5 | 97.5 | 93.8 - 100 | 100 | 99 | 99.5 | 100 | 100 | 0.5 | 99.5 | 98.8 - 100 | 100 | 92.6 | 99.5 | 100 | 100 | 0.5 | 98.2 | 95.7 - 100 |
| | PB – left | 99.8 | 84.1 | 98.4 | 100 | 100 | 1.6 | 95.6 | 90.5 - 100 | 99.8 | 83 | 97.7 | 100 | 100 | 2.3 | 95.2 | 90.1 - 100 | 99.5 | 83.5 | 98.2 | 100 | 100 | 1.8 | 95.4 | 90.3 - 100 |
| | PB – right | 100 | 90.2 | 99 | 100 | 100 | 1 | 97.6 | 95 - 100 | 100 | 84.8 | 99 | 100 | 100 | 1 | 96.6 | 93.7 - 99.6 | 100 | 87.8 | 99 | 100 | 100 | 1 | 97.1 | 94.4 - 99.8 |
| | PF – left | 100 | 67.8 | 99 | 100 | 100 | 1 | 96.3 | 91.9 - 100 | 100 | 69.3 | 98.9 | 100 | 100 | 1.1 | 96.3 | 91.9 - 100 | 100 | 68.6 | 99.2 | 100 | 100 | 0.8 | 96.3 | 91.9 - 100 |
| | PF – right | 100 | 84.8 | 99.3 | 100 | 100 | 0.7 | 97.8 | 95.8 - 99.9 | 100 | 84.4 | 98.7 | 100 | 100 | 1.3 | 97.4 | 95.1 - 99.8 | 100 | 84.6 | 98.7 | 100 | 100 | 1.3 | 97.6 | 95.4 - 99.9 |
| All PwMS | BB | 100 | 95.8 | 99.5 | 100 | 100 | 0.5 | 98.2 | 95.7 - 100 | 100 | 96.5 | 100 | 100 | 100 | 0 | 98.1 | 95.6 - 100 | 100 | 96.8 | 99.5 | 100 | 100 | 0.5 | 98.2 | 95.7 - 100 |
| | BF | 100 | 97.5 | 100 | 100 | 100 | 0 | 99 | 97.9 - 100 | 100 | 98.9 | 100 | 100 | 100 | 0 | 99.7 | 99.5 - 100 | 100 | 98.3 | 99.8 | 100 | 100 | 0.2 | 99.3 | 98.5 - 100 |
| | PB – left | 100 | 92.4 | 99.1 | 100 | 100 | 0.9 | 98.3 | 96.7 - 99.9 | 100 | 90.2 | 100 | 100 | 100 | 0 | 97.7 | 95.9 - 99.5 | 100 | 90.8 | 99.3 | 100 | 100 | 0.7 | 98 | 96.3 - 99.6 |
| | PB – right | 100 | 95.8 | 99.6 | 100 | 100 | 0.4 | 99 | 98.2 - 99.8 | 100 | 91.4 | 99.5 | 100 | 100 | 0.5 | 98.6 | 97.7 - 99.6 | 100 | 95.2 | 99.3 | 100 | 100 | 0.7 | 98.8 | 97.9 - 99.7 |
| | PF – left | 100 | 97.9 | 99.8 | 100 | 100 | 0.2 | 98.2 | 96.5 - 99.9 | 100 | 97.9 | 100 | 100 | 100 | 0 | 98.2 | 96.5 - 100 | 100 | 98 | 99.6 | 100 | 100 | 0.4 | 98.2 | 96.5 - 100 |
| | PF – right | 100 | 97.5 | 100 | 100 | 100 | 0 | 99.1 | 98.5 - 99.8 | 100 | 97.4 | 100 | 100 | 100 | 0 | 99.1 | 98.4 - 99.9 | 100 | 98.2 | 99.7 | 100 | 100 | 0.3 | 99.1 | 98.4 - 99.8 |
| ***Structured in-lab overground 2MWT – smartphone vs Gait Up*** | | | | | | | | | | | | | | | | | | | | | | | | | |
| HC | BB | 100 | 97.8 | 99.6 | 100 | 100 | 0.4 | 99.6 | 99.3 - 99.9 | 99.4 | 97.2 | 98.5 | 99.9 | 100 | 1.4 | 99 | 98.6 - 99.3 | 99.5 | 98.2 | 98.7 | 99.7 | 100 | 1 | 99.3 | 99 - 99.5 |
| | BF | 100 | 97 | 99.5 | 100 | 100 | 0.5 | 99.5 | 99.1 - 99.8 | 99.5 | 97.5 | 98.9 | 100 | 100 | 1.1 | 99.3 | 99 - 99.6 | 99.7 | 97.9 | 99 | 100 | 100 | 1 | 99.4 | 99.1 - 99.6 |
| | PB – left | 99.8 | 96.7 | 99.4 | 100 | 100 | 0.6 | 99.3 | 98.9 - 99.8 | 99.2 | 95.5 | 98.3 | 99.5 | 100 | 1.2 | 98.6 | 98 - 99.2 | 99.2 | 96.6 | 98.7 | 99.7 | 100 | 1 | 98.9 | 98.6 - 99.3 |
| | PB – right | 100 | 96.5 | 99.2 | 100 | 100 | 0.8 | 99.2 | 98.7 - 99.7 | 99.4 | 97 | 98.5 | 99.9 | 100 | 1.4 | 98.9 | 98.5 - 99.3 | 99.4 | 97.1 | 98.7 | 99.7 | 100 | 1.1 | 99 | 98.7 - 99.4 |
| | PF – left | 99.4 | 95.1 | 98.3 | 100 | 100 | 1.7 | 98.7 | 98 - 99.4 | 98.7 | 96.8 | 97.9 | 99.5 | 100 | 1.6 | 98.7 | 98.3 - 99.1 | 99 | 96.5 | 98.1 | 99.7 | 100 | 1.6 | 98.7 | 98.3 - 99.1 |
| | PF – right | 99.8 | 95.4 | 99.1 | 100 | 100 | 0.9 | 99 | 98.4 - 99.6 | 99.8 | 96.5 | 98.9 | 100 | 100 | 1.1 | 99.2 | 98.8 - 99.6 | 99.7 | 96.6 | 98.4 | 99.9 | 100 | 1.6 | 99.1 | 98.7 - 99.5 |
| Mild PwMS | BB | 100 | 98.4 | 100 | 100 | 100 | 0 | 99.7 | 99.4 - 100 | 99.4 | 97.2 | 97.7 | 100 | 100 | 2.3 | 98.9 | 98.4 - 99.4 | 99.7 | 98.3 | 98.7 | 100 | 100 | 1.3 | 99.3 | 99 - 99.6 |
| | BF | 100 | 98.8 | 99.5 | 100 | 100 | 0.5 | 99.6 | 99.2 - 100 | 100 | 97.7 | 99.1 | 100 | 100 | 0.9 | 99.4 | 99 - 99.8 | 99.7 | 98.3 | 99.3 | 100 | 100 | 0.7 | 99.5 | 99.1 - 99.9 |
| | PB – left | 100 | 97.4 | 98.9 | 100 | 100 | 1.1 | 99.2 | 98.5 - 99.9 | 98.8 | 95 | 97.7 | 100 | 100 | 2.3 | 98.4 | 97.7 - 99.1 | 99.1 | 97.2 | 98.1 | 99.7 | 100 | 1.7 | 98.8 | 98.3 - 99.3 |
| | PB – right | 100 | 97.4 | 98.8 | 100 | 100 | 1.2 | 99 | 98.3 - 99.7 | 97.8 | 96.3 | 97 | 100 | 100 | 3 | 98.2 | 97.5 - 98.9 | 98.9 | 97.2 | 97.6 | 99.7 | 100 | 2.1 | 98.6 | 98 - 99.1 |
| | PF – left | 99 | 97.7 | 98.8 | 100 | 100 | 1.2 | 98.9 | 98.2 - 99.6 | 99.4 | 93.2 | 96.7 | 100 | 100 | 3.3 | 98.2 | 97.2 - 99.2 | 99.2 | 96.3 | 97.8 | 99.7 | 100 | 1.9 | 98.5 | 97.8 - 99.3 |
| | PF – right | 99.4 | 93.6 | 98.3 | 100 | 100 | 1.7 | 98.6 | 97.7 - 99.5 | 99 | 96.4 | 98.3 | 99.5 | 100 | 1.2 | 98.7 | 98.3 - 99.2 | 98.9 | 96.2 | 98.2 | 99.5 | 99.8 | 1.3 | 98.6 | 98.1 - 99.2 |
| Moderate PwMS | BB | 100 | 99.4 | 100 | 100 | 100 | 0 | 99.9 | 99.8 - 100 | 99 | 97.2 | 98.4 | 100 | 100 | 1.6 | 99 | 98.6 - 99.3 | 99.5 | 98.4 | 99.2 | 100 | 100 | 0.8 | 99.4 | 99.2 - 99.7 |
| | BF | 100 | 99 | 100 | 100 | 100 | 0 | 99.6 | 99.1 - 100 | 100 | 95.1 | 98.8 | 100 | 100 | 1.2 | 99 | 98.3 - 99.7 | 100 | 97.2 | 99.4 | 100 | 100 | 0.6 | 99.3 | 98.7 - 99.9 |
| | PB – left | 100 | 99.1 | 100 | 100 | 100 | 0 | 99.9 | 99.8 - 100 | 98.7 | 95.6 | 97.8 | 99.4 | 100 | 1.6 | 98.4 | 97.9 - 98.8 | 99.3 | 97.8 | 98.8 | 99.7 | 100 | 0.9 | 99.1 | 98.9 - 99.4 |
| | PB – right | 100 | 99 | 100 | 100 | 100 | 0 | 99.5 | 98.7 - 100 | 98.6 | 94.8 | 97.4 | 98.9 | 100 | 1.5 | 97.7 | 96.5 - 98.9 | 99.2 | 94.7 | 98.6 | 99.5 | 100 | 0.9 | 98.6 | 97.7 - 99.4 |
| | PF – left | 100 | 98.9 | 100 | 100 | 100 | 0 | 99.8 | 99.7 - 100 | 99.4 | 96.9 | 98.4 | 100 | 100 | 1.6 | 99 | 98.7 - 99.4 | 99.7 | 97.9 | 99.2 | 100 | 100 | 0.8 | 99.4 | 99.2 - 99.7 |
| | PF – right | 100 | 98.5 | 100 | 100 | 100 | 0 | 99.7 | 99.5 - 99.9 | 99.4 | 96.9 | 98.8 | 100 | 100 | 1.2 | 99 | 98.5 - 99.5 | 99.5 | 97.9 | 99.4 | 100 | 100 | 0.6 | 99.4 | 99.1 - 99.7 |
| Severe PwMS | BB | 100 | 97.2 | 100 | 100 | 100 | 0 | 97.4 | 93 - 100 | 99.3 | 97.8 | 98.6 | 100 | 100 | 1.4 | 97 | 92.4 - 100 | 99.6 | 98.1 | 99.3 | 100 | 100 | 0.7 | 97.2 | 92.7 - 100 |
| | BF | 100 | 98.4 | 99.9 | 100 | 100 | 0.1 | 98.4 | 95.8 - 100 | 99.4 | 97.8 | 99.2 | 100 | 100 | 0.8 | 98.3 | 96.1 - 100 | 99.7 | 98.8 | 99.3 | 100 | 100 | 0.7 | 98.4 | 95.9 - 100 |
| | PB – left | 100 | 98.2 | 99.3 | 100 | 100 | 0.7 | 99.4 | 98.8 - 100 | 98.6 | 95.9 | 98.1 | 100 | 100 | 1.9 | 98.6 | 98 - 99.1 | 99.3 | 97.5 | 98.8 | 100 | 100 | 1.2 | 99 | 98.4 - 99.5 |
| | PB – right | 100 | 94.7 | 99.3 | 100 | 100 | 0.7 | 98.9 | 97.8 - 100 | 98.5 | 92.4 | 97.5 | 99.1 | 100 | 1.7 | 97.7 | 96.9 - 98.5 | 99 | 93.9 | 98.6 | 99.6 | 100 | 0.9 | 98.3 | 97.4 - 99.2 |
| | PF – left | 100 | 97.9 | 99.4 | 100 | 100 | 0.6 | 99.4 | 98.9 - 100 | 99.4 | 94.5 | 98.6 | 100 | 100 | 1.4 | 98.7 | 98.1 - 99.4 | 99.7 | 95.6 | 98.8 | 100 | 100 | 1.2 | 99.1 | 98.6 - 99.6 |

| | PF – right | 100 | 89 | 99.2 | 100 | 100 | 0.8 | 98.4 | 96.6 - 100 | 98.8 | 90.6 | 97.6 | 99.4 | 100 | 1.8 | 97.6 | 96.2 - 99 | 99.3 | 88.8 | 98.4 | 99.7 | 100 | 1.2 | 98 | 96.4 - 99.5 |
|---|---|---|---|---|---|---|---|---|---|---|---|---|---|---|---|---|---|---|---|---|---|---|---|---|---|
| All PwMS | BB | 100 | 98.6 | 100 | 100 | 100 | 0 | 98.9 | 97.3 - 100 | 99.3 | 97.2 | 98.4 | 100 | 100 | 1.6 | 98.2 | 96.5 - 99.9 | 99.6 | 98.3 | 98.9 | 100 | 100 | 1.1 | 98.6 | 96.9 - 100 |
| | BF | 100 | 98.6 | 100 | 100 | 100 | 0 | 99.2 | 98.2 - 100 | 100 | 96.8 | 99.1 | 100 | 100 | 0.9 | 98.8 | 98 - 99.7 | 99.7 | 98.2 | 99.3 | 100 | 100 | 0.7 | 99 | 98.1 - 99.9 |
| | PB – left | 100 | 98 | 99.6 | 100 | 100 | 0.4 | 99.5 | 99.3 - 99.8 | 98.7 | 95.1 | 97.9 | 99.5 | 100 | 1.5 | 98.4 | 98.1 - 98.8 | 99.3 | 97.2 | 98.8 | 99.7 | 100 | 0.9 | 99 | 98.7 - 99.2 |
| | PB – right | 100 | 96.2 | 99.4 | 100 | 100 | 0.6 | 99.2 | 98.6 - 99.7 | 98.4 | 93.6 | 97.3 | 99.3 | 100 | 2.1 | 97.8 | 97.2 - 98.4 | 99 | 94.4 | 98.5 | 99.6 | 100 | 1.1 | 98.5 | 98 - 99 |
| | PF – left | 100 | 98 | 99.1 | 100 | 100 | 0.9 | 99.5 | 99.2 - 99.7 | 99.4 | 94.9 | 98.3 | 100 | 100 | 1.7 | 98.7 | 98.3 - 99.1 | 99.7 | 96.4 | 98.8 | 100 | 100 | 1.2 | 99.1 | 98.8 - 99.4 |
| | PF – right | 100 | 95.3 | 99.2 | 100 | 100 | 0.8 | 98.9 | 98.2 - 99.6 | 99.1 | 95.6 | 98.1 | 100 | 100 | 1.9 | 98.4 | 97.8 - 99 | 99.4 | 95.8 | 98.8 | 99.7 | 100 | 0.9 | 98.7 | 98 - 99.3 |
| ***Structured real-world walking – smartphone vs Gait Up*** | | | | | | | | | | | | | | | | | | | | | | | | | |
| HC | BF suggested location | 100 | 99.8 | 100 | 100 | 100 | 0 | 99.9 | 99.9 - 100 | 100 | 98.8 | 100 | 100 | 100 | 0 | 99.8 | 99.7 - 100 | 100 | 99.3 | 99.9 | 100 | 100 | 0.2 | 99.9 | 99.8 - 100 |
| Mild PwMS | BF suggested location | 100 | 99.9 | 100 | 100 | 100 | 0 | 99.9 | 99.8 - 100 | 100 | 98.1 | 100 | 100 | 100 | 0 | 99.2 | 97.7 - 100 | 100 | 98.9 | 100 | 100 | 100 | 0 | 99.5 | 98.7 - 100 |
| Moderate PwMS | BF suggested location | 100 | 98.3 | 99.7 | 100 | 100 | 0 | 99.5 | 99 - 100 | 99.9 | 93.3 | 98.4 | 100 | 100 | 1.2 | 98.7 | 97.9 - 99.5 | 99.8 | 95.9 | 99.1 | 100 | 100 | 0.7 | 99.1 | 98.5 - 99.7 |
| Severe PwMS | BF suggested location | 100 | 79.5 | 98 | 100 | 100 | 2.8 | 95.9 | 92.9 - 98.9 | 98.9 | 80.3 | 96.4 | 100 | 100 | 5.4 | 95.4 | 92.5 - 98.4 | 99.3 | 79.9 | 97.4 | 100 | 100 | 4.5 | 95.7 | 92.7 - 98.7 |
| All PwMS | BF suggested location | 100 | 89 | 99.6 | 100 | 100 | 0.5 | 98.3 | 97.1 - 99.5 | 100 | 85.9 | 98.4 | 100 | 100 | 1.5 | 97.6 | 96.4 - 98.9 | 99.9 | 88.8 | 99 | 100 | 100 | 1 | 97.9 | 96.7 - 99.1 |
| ***Unstructured real-world walking– smartphone vs Gait Up*** | | | | | | | | | | | | | | | | | | | | | | | | | |
| HC | User-preferred location | 97.4 | 92.7 | 94.8 | 98.4 | 99.2 | 5.1 | 96.5 | 95.6 - 97.4 | 98.7 | 94.8 | 97.4 | 99.5 | 99.8 | 3.2 | 98.1 | 97.4 - 98.8 | 97.8 | 92.6 | 96.2 | 98.8 | 99.4 | 4.5 | 97.1 | 96.3 - 97.9 |
| Mild PwMS | User-preferred location | 95.2 | 87 | 91.2 | 97.7 | 99.6 | 5.8 | 94 | 91.7 - 96.3 | 97.8 | 94.3 | 95.5 | 98.5 | 99.8 | 3.8 | 97.3 | 96.5 - 98.1 | 96 | 88.8 | 94 | 97.6 | 99.7 | 4.4 | 95.3 | 93.7 - 96.9 |
| Moderate PwMS | User-preferred location | 93.4 | 80.6 | 91.7 | 95.4 | 99 | 6.5 | 92.1 | 89.8 - 94.3 | 96.9 | 76 | 94.2 | 98.4 | 99.8 | 5.8 | 94.4 | 91.7 - 97.1 | 94.8 | 76.1 | 92.6 | 96.8 | 99.1 | 5.9 | 93 | 90.6 - 95.4 |
| Severe PwMS | User-preferred location | 89.6 | 73 | 83.3 | 93.1 | 97.1 | 11.4 | 88.1 | 85.5 - 90.8 | 93.9 | 80.9 | 90.3 | 96.8 | 99.1 | 9.3 | 92.3 | 90 - 94.6 | 91.8 | 78.5 | 86.6 | 94.7 | 97.4 | 10.1 | 89.9 | 87.6 - 92.2 |
| All PwMS | User-preferred location | 93 | 77.5 | 88.9 | 95.9 | 98.8 | 8 | 91.1 | 89.6 - 92.6 | 96.4 | 80.9 | 93.4 | 98.4 | 99.7 | 6.5 | 94.3 | 92.9 - 95.7 | 94.4 | 79 | 90.4 | 96.6 | 99 | 7.1 | 92.4 | 91 - 93.8 |

[a]Mild PwMS are PwMS with mild EDDS scores (EDSS: ≤ 3.5), moderate PwMS are PwMS with moderate EDSS scores (EDSS: 4.0 - 5.5), and severe PwMS are PwMS with severe EDSS scores (EDSS: 5.5 - 6.5).

BB = belt back; BF = belt front; CI = confidence interval; EDSS = Expanded Disability Status Scale; IQR = within-subject interquartile range; Mdn = median; P05 = 5th percentile; P95 = 95th percentile; PB = pocket back; PF = pocket front; PwMS = people with multiple sclerosis; Q1 = first quantile; Q3: = third quantile; ws-IQR = within-subject interquartile range.

**Supplementary Table S2. Algorithm performance for final contacts.**

| Cohort[a] | Smartphone location | Precision | | | | | | | | Recall | | | | | | | | F1 score | | | | | | | |
|---|---|---|---|---|---|---|---|---|---|---|---|---|---|---|---|---|---|---|---|---|---|---|---|---|---|
| | | Mdn | P05 | Q1 | Q3 | P95 | ws-IQR | Mean | 95% CI | Mdn | P05 | Q1 | Q3 | P95 | ws-IQR | Mean | 95% CI | Mdn | P05 | Q1 | Q3 | P95 | ws-IQR | Mean | 95% CI |
| ***Structured in-lab treadmill 2MWT – smartphone vs mocap*** | | | | | | | | | | | | | | | | | | | | | | | | | |
| HC | BB | 100 | 100 | 100 | 100 | 100 | 0 | 100 | 99.9 - 100 | 100 | 100 | 100 | 100 | 100 | 0 | 100 | 100 - 100 | 100 | 100 | 100 | 100 | 100 | 0 | 100 | 99.9 - 100 |
| | BF | 100 | 100 | 100 | 100 | 100 | 0 | 100 | 99.9 - 100 | 100 | 100 | 100 | 100 | 100 | 0 | 100 | 100 - 100 | 100 | 100 | 100 | 100 | 100 | 0 | 100 | 99.9 - 100 |
| | PB – left | 100 | 99.9 | 100 | 100 | 100 | 0 | 99.9 | 99.9 - 100 | 100 | 99.2 | 100 | 100 | 100 | 0 | 99.8 | 99.5 - 100 | 100 | 99.4 | 100 | 100 | 100 | 0 | 99.9 | 99.7 - 100 |
| | PB – right | 100 | 99.5 | 100 | 100 | 100 | 0 | 99.9 | 99.8 - 100 | 100 | 99.9 | 100 | 100 | 100 | 0 | 100 | 99.9 - 100 | 100 | 99.6 | 100 | 100 | 100 | 0 | 99.9 | 99.9 - 100 |
| | PF – left | 100 | 100 | 100 | 100 | 100 | 0 | 100 | 99.9 - 100 | 100 | 99.9 | 100 | 100 | 100 | 0 | 100 | 99.9 - 100 | 100 | 99.7 | 100 | 100 | 100 | 0 | 100 | 99.9 - 100 |
| | PF – right | 100 | 100 | 100 | 100 | 100 | 0 | 100 | 99.9 - 100 | 100 | 99.4 | 100 | 100 | 100 | 0 | 99.9 | 99.8 - 100 | 100 | 99.5 | 100 | 100 | 100 | 0 | 99.9 | 99.9 - 100 |
| Mild PwMS | BB | 100 | 100 | 100 | 100 | 100 | 0 | 100 | 100 - 100 | 100 | 100 | 100 | 100 | 100 | 0 | 100 | 99.9 - 100 | 100 | 100 | 100 | 100 | 100 | 0 | 100 | 99.9 - 100 |
| | BF | 100 | 99.6 | 100 | 100 | 100 | 0 | 100 | 99.9 - 100 | 100 | 99.3 | 100 | 100 | 100 | 0 | 99.9 | 99.8 - 100 | 100 | 99.7 | 100 | 100 | 100 | 0 | 99.9 | 99.9 - 100 |
| | PB – left | 100 | 100 | 100 | 100 | 100 | 0 | 99.9 | 99.8 - 100 | 100 | 98.9 | 100 | 100 | 100 | 0 | 99.8 | 99.7 - 100 | 100 | 99.1 | 100 | 100 | 100 | 0 | 99.9 | 99.8 - 100 |
| | PB – right | 100 | 100 | 100 | 100 | 100 | 0 | 99.9 | 99.9 - 100 | 100 | 99.3 | 100 | 100 | 100 | 0 | 99.9 | 99.7 - 100 | 100 | 99.1 | 100 | 100 | 100 | 0 | 99.9 | 99.8 - 100 |
| | PF – left | 100 | 100 | 100 | 100 | 100 | 0 | 100 | 99.9 - 100 | 100 | 98.6 | 100 | 100 | 100 | 0 | 99.6 | 99.1 - 100 | 100 | 99.3 | 99.9 | 100 | 100 | 0.1 | 99.8 | 99.5 - 100 |
| | PF – right | 100 | 100 | 100 | 100 | 100 | 0 | 100 | 100 - 100 | 100 | 99.7 | 100 | 100 | 100 | 0 | 100 | 99.9 - 100 | 100 | 99.8 | 100 | 100 | 100 | 0 | 100 | 100 - 100 |
| Moderate PwMS | BB | 100 | 99.5 | 100 | 100 | 100 | 0 | 99.9 | 99.8 - 100 | 100 | 99.5 | 100 | 100 | 100 | 0 | 99.9 | 99.6 - 100 | 100 | 99.4 | 100 | 100 | 100 | 0 | 99.9 | 99.8 - 100 |
| | BF | 100 | 98.9 | 100 | 100 | 100 | 0 | 99.9 | 99.7 - 100 | 100 | 98.7 | 100 | 100 | 100 | 0 | 99.7 | 99.4 - 100 | 100 | 98.9 | 100 | 100 | 100 | 0 | 99.8 | 99.6 - 99.9 |
| | PB – left | 100 | 99.5 | 100 | 100 | 100 | 0 | 99.9 | 99.8 - 100 | 100 | 93.5 | 100 | 100 | 100 | 0 | 98.2 | 96.1 - 100 | 100 | 96.5 | 100 | 100 | 100 | 0 | 98.9 | 97.6 - 100 |
| | PB – right | 100 | 99.1 | 100 | 100 | 100 | 0 | 99.9 | 99.8 - 100 | 100 | 94.7 | 99.6 | 100 | 100 | 0.4 | 99.3 | 98.8 - 99.9 | 100 | 97.3 | 99.8 | 100 | 100 | 0.2 | 99.6 | 99.3 - 99.9 |
| | PF – left | 100 | 99.5 | 100 | 100 | 100 | 0 | 99.3 | 98 - 100 | 100 | 97.2 | 99.8 | 100 | 100 | 0.2 | 98.9 | 97.5 - 100 | 100 | 98.6 | 99.8 | 100 | 100 | 0.2 | 99.1 | 97.7 - 100 |
| | PF – right | 100 | 98.9 | 100 | 100 | 100 | 0 | 99.9 | 99.8 - 100 | 100 | 98.9 | 100 | 100 | 100 | 0 | 99.8 | 99.6 - 100 | 100 | 99.1 | 100 | 100 | 100 | 0 | 99.8 | 99.7 - 100 |
| Severe PwMS | BB | 100 | 91 | 99.5 | 100 | 100 | 0.5 | 95.1 | 86.9 - 100 | 100 | 80.2 | 99 | 100 | 100 | 1 | 94.2 | 86 - 100 | 100 | 85.2 | 99.5 | 100 | 100 | 0.5 | 94.6 | 86.5 - 100 |
| | BF | 100 | 62.3 | 99.5 | 100 | 100 | 0.5 | 93.7 | 84.9 - 100 | 100 | 95.3 | 99.4 | 100 | 100 | 0.6 | 95.3 | 87.2 - 100 | 100 | 75.7 | 99.3 | 100 | 100 | 0.7 | 94.2 | 85.9 - 100 |
| | PB – left | 100 | 13.4 | 99.5 | 100 | 100 | 0.5 | 90.8 | 79.6 - 100 | 100 | 12.6 | 98.5 | 100 | 100 | 1.5 | 90.1 | 78.9 - 100 | 100 | 13 | 99.2 | 100 | 100 | 0.8 | 90.4 | 79.2 - 100 |
| | PB – right | 100 | 97.3 | 99.8 | 100 | 100 | 0.2 | 99.5 | 99 - 99.9 | 100 | 90.5 | 98.9 | 100 | 100 | 1.1 | 97.8 | 96 - 99.6 | 99.9 | 95 | 99 | 100 | 100 | 1 | 98.6 | 97.5 - 99.7 |
| | PF – left | 100 | 98.2 | 99.4 | 100 | 100 | 0.6 | 98.1 | 94.9 - 100 | 99.6 | 92.6 | 99.3 | 100 | 100 | 0.7 | 96.8 | 92.6 - 100 | 99.8 | 95.9 | 99.3 | 100 | 100 | 0.7 | 97.4 | 93.6 - 100 |
| | PF – right | 100 | 97 | 99.4 | 100 | 100 | 0.6 | 99.2 | 98.3 - 100 | 100 | 90.4 | 97.5 | 100 | 100 | 2.5 | 98 | 96.6 - 99.4 | 99.7 | 93.6 | 98.4 | 100 | 100 | 1.6 | 98.6 | 97.5 - 99.7 |
| All PwMS | BB | 100 | 99.1 | 100 | 100 | 100 | 0 | 98.5 | 96.1 - 100 | 100 | 96.3 | 100 | 100 | 100 | 0 | 98.2 | 95.8 - 100 | 100 | 97.8 | 100 | 100 | 100 | 0 | 98.4 | 96 - 100 |
| | BF | 100 | 98.8 | 100 | 100 | 100 | 0 | 98.1 | 95.5 - 100 | 100 | 98.2 | 100 | 100 | 100 | 0 | 98.5 | 96.1 - 100 | 100 | 98.8 | 99.7 | 100 | 100 | 0.3 | 98.2 | 95.7 - 100 |
| | PB – left | 100 | 98.3 | 100 | 100 | 100 | 0 | 97.3 | 93.9 - 100 | 100 | 86.7 | 99.8 | 100 | 100 | 0.2 | 96.3 | 92.8 - 99.8 | 100 | 89.4 | 99.8 | 100 | 100 | 0.2 | 96.7 | 93.3 - 100 |
| | PB – right | 100 | 98.4 | 100 | 100 | 100 | 0 | 99.8 | 99.6 - 99.9 | 100 | 94.2 | 99.6 | 100 | 100 | 0.4 | 99 | 98.4 - 99.7 | 100 | 95.7 | 99.8 | 100 | 100 | 0.2 | 99.4 | 99 - 99.8 |
| | PF – left | 100 | 98.8 | 100 | 100 | 100 | 0 | 99.1 | 98.1 - 100 | 100 | 95.4 | 99.5 | 100 | 100 | 0.5 | 98.5 | 97.1 - 99.9 | 100 | 97.6 | 99.6 | 100 | 100 | 0.4 | 98.8 | 97.6 - 100 |
| | PF – right | 100 | 98.9 | 100 | 100 | 100 | 0 | 99.7 | 99.4 - 100 | 100 | 96.4 | 100 | 100 | 100 | 0 | 99.3 | 98.9 - 99.8 | 100 | 97.9 | 99.8 | 100 | 100 | 0.2 | 99.5 | 99.2 - 99.9 |
| ***Structured in-lab treadmill 2MWT – smartphone vs Gait Up*** | | | | | | | | | | | | | | | | | | | | | | | | | |
| HC | BB | 100 | 100 | 100 | 100 | 100 | 0 | 100 | 100 - 100 | 100 | 100 | 100 | 100 | 100 | 0 | 100 | 100 - 100 | 100 | 100 | 100 | 100 | 100 | 0 | 100 | 100 - 100 |
| | BF | 100 | 100 | 100 | 100 | 100 | 0 | 100 | 100 - 100 | 100 | 100 | 100 | 100 | 100 | 0 | 100 | 100 - 100 | 100 | 100 | 100 | 100 | 100 | 0 | 100 | 100 - 100 |
| | PB – left | 100 | 99.9 | 100 | 100 | 100 | 0 | 100 | 99.9 - 100 | 100 | 99.2 | 100 | 100 | 100 | 0 | 99.8 | 99.5 - 100 | 100 | 99.5 | 100 | 100 | 100 | 0 | 99.9 | 99.7 - 100 |
| | PB – right | 100 | 99.9 | 100 | 100 | 100 | 0 | 100 | 99.9 - 100 | 100 | 100 | 100 | 100 | 100 | 0 | 100 | 99.9 - 100 | 100 | 99.8 | 100 | 100 | 100 | 0 | 100 | 99.9 - 100 |
| | PF – left | 100 | 100 | 100 | 100 | 100 | 0 | 100 | 100 - 100 | 100 | 99.8 | 100 | 100 | 100 | 0 | 100 | 99.9 - 100 | 100 | 99.8 | 100 | 100 | 100 | 0 | 100 | 100 - 100 |
| | PF – right | 100 | 100 | 100 | 100 | 100 | 0 | 100 | 100 - 100 | 100 | 99.8 | 100 | 100 | 100 | 0 | 99.9 | 99.9 - 100 | 100 | 99.9 | 100 | 100 | 100 | 0 | 100 | 99.9 - 100 |
| Mild PwMS | BB | 100 | 99.6 | 100 | 100 | 100 | 0 | 99.9 | 99.9 - 100 | 100 | 99.6 | 100 | 100 | 100 | 0 | 99.9 | 99.8 - 100 | 100 | 99.6 | 100 | 100 | 100 | 0 | 99.9 | 99.9 - 100 |
| | BF | 100 | 99.6 | 100 | 100 | 100 | 0 | 99.9 | 99.9 - 100 | 100 | 99.3 | 100 | 100 | 100 | 0 | 99.9 | 99.7 - 100 | 100 | 99.6 | 100 | 100 | 100 | 0 | 99.9 | 99.8 - 100 |
| | PB – left | 100 | 99.6 | 100 | 100 | 100 | 0 | 99.9 | 99.8 - 100 | 100 | 98.8 | 100 | 100 | 100 | 0 | 99.8 | 99.6 - 100 | 100 | 99 | 100 | 100 | 100 | 0 | 99.9 | 99.7 - 100 |
| | PB – right | 100 | 99.6 | 100 | 100 | 100 | 0 | 99.9 | 99.8 - 100 | 100 | 99.2 | 100 | 100 | 100 | 0 | 99.8 | 99.6 - 100 | 100 | 99 | 100 | 100 | 100 | 0 | 99.9 | 99.8 - 100 |
| | PF – left | 100 | 99.6 | 100 | 100 | 100 | 0 | 99.9 | 99.9 - 100 | 100 | 99 | 100 | 100 | 100 | 0 | 99.6 | 99.1 - 100 | 100 | 99.5 | 100 | 100 | 100 | 0 | 99.8 | 99.5 - 100 |
| | PF – right | 100 | 99.6 | 100 | 100 | 100 | 0 | 100 | 99.9 - 100 | 100 | 99.7 | 100 | 100 | 100 | 0 | 99.9 | 99.9 - 100 | 100 | 99.8 | 100 | 100 | 100 | 0 | 100 | 99.9 - 100 |
| Moderate PwMS | BB | 100 | 98.2 | 100 | 100 | 100 | 0 | 99.7 | 99.5 - 99.9 | 100 | 98.7 | 100 | 100 | 100 | 0 | 99.8 | 99.5 - 100 | 100 | 98.2 | 100 | 100 | 100 | 0 | 99.7 | 99.5 - 100 |
| | BF | 100 | 97.8 | 100 | 100 | 100 | 0 | 99.6 | 99.4 - 99.9 | 100 | 98 | 100 | 100 | 100 | 0 | 99.7 | 99.4 - 100 | 100 | 98.1 | 100 | 100 | 100 | 0 | 99.7 | 99.4 - 100 |
| | PB – left | 100 | 98.2 | 100 | 100 | 100 | 0 | 99.6 | 99.4 - 99.9 | 100 | 92.9 | 99.6 | 100 | 100 | 0.4 | 98.2 | 96.1 - 100 | 100 | 95.9 | 99.2 | 100 | 100 | 0.8 | 98.8 | 97.5 - 100 |
| | PB – right | 100 | 97.9 | 100 | 100 | 100 | 0 | 99.7 | 99.4 - 99.9 | 100 | 94.6 | 99.6 | 100 | 100 | 0.4 | 99.2 | 98.6 - 99.8 | 100 | 96.9 | 99.5 | 100 | 100 | 0.5 | 99.4 | 99.1 - 99.8 |
| | PF – left | 100 | 97.6 | 100 | 100 | 100 | 0 | 98.6 | 96.2 - 100 | 100 | 97 | 99.7 | 100 | 100 | 0.3 | 98.4 | 95.9 - 100 | 100 | 98 | 99.6 | 100 | 100 | 0.4 | 98.5 | 96.1 - 100 |
| | PF – right | 100 | 98.2 | 100 | 100 | 100 | 0 | 99.7 | 99.5 - 99.9 | 100 | 98.3 | 100 | 100 | 100 | 0 | 99.7 | 99.4 - 100 | 100 | 98.5 | 99.6 | 100 | 100 | 0.4 | 99.7 | 99.5 - 99.9 |
| Severe PwMS | BB | 100 | 91.9 | 99 | 100 | 100 | 1 | 95 | 86.9 - 100 | 100 | 81 | 98.4 | 100 | 100 | 1.6 | 94.3 | 86.1 - 100 | 100 | 86.1 | 98.3 | 100 | 100 | 1.7 | 94.6 | 86.5 - 100 |
| | BF | 100 | 62.6 | 99.2 | 100 | 100 | 0.8 | 93.8 | 85.4 - 100 | 100 | 97.4 | 99.4 | 100 | 100 | 0.6 | 95.7 | 87.9 - 100 | 100 | 76 | 99.2 | 100 | 100 | 0.8 | 94.5 | 86.5 - 100 |
| | PB – left | 100 | 23.9 | 99.3 | 100 | 100 | 0.7 | 91.6 | 81.3 - 100 | 99.7 | 23.1 | 97.5 | 100 | 100 | 2.5 | 90.9 | 80.6 - 100 | 99.7 | 23.5 | 98.1 | 100 | 100 | 1.9 | 91.3 | 81 - 100 |
| | PB – right | 100 | 97.7 | 99.3 | 100 | 100 | 0.7 | 99.5 | 99.1 - 99.8 | 99.8 | 91.4 | 98.1 | 100 | 100 | 1.9 | 98 | 96.3 - 99.6 | 99.8 | 94.9 | 98.7 | 100 | 100 | 1.3 | 98.7 | 97.7 - 99.6 |
| | PF – left | 100 | 85.4 | 99.3 | 100 | 100 | 0.7 | 97.5 | 95 - 99.9 | 99.8 | 79.4 | 99.2 | 100 | 100 | 0.8 | 96.3 | 92.4 - 100 | 99.8 | 82.3 | 98.5 | 100 | 100 | 1.5 | 96.8 | 93.6 - 100 |
| | PF – right | 100 | 96.6 | 98.7 | 100 | 100 | 1.3 | 97.5 | 93.7 - 100 | 100 | 90.6 | 97.8 | 100 | 100 | 2.2 | 96.6 | 92.6 - 100 | 99.7 | 93.5 | 98.2 | 100 | 100 | 1.8 | 97 | 93.1 - 100 |
| All PwMS | BB | 100 | 97.2 | 100 | 100 | 100 | 0 | 98.3 | 95.9 - 100 | 100 | 96.2 | 100 | 100 | 100 | 0 | 98.1 | 95.6 - 100 | 100 | 97.7 | 99.7 | 100 | 100 | 0.3 | 98.2 | 95.8 - 100 |
| | BF | 100 | 97.4 | 99.8 | 100 | 100 | 0.2 | 98 | 95.4 - 100 | 100 | 98.6 | 100 | 100 | 100 | 0 | 98.5 | 96.2 - 100 | 100 | 97.5 | 99.7 | 100 | 100 | 0.3 | 98.2 | 95.7 - 100 |
| | PB – left | 100 | 97.5 | 100 | 100 | 100 | 0 | 97.3 | 94.1 - 100 | 100 | 90.8 | 99.3 | 100 | 100 | 0.7 | 96.4 | 93.1 - 99.7 | 100 | 94.2 | 99.2 | 100 | 100 | 0.8 | 96.8 | 93.6 - 100 |
| | PB – right | 100 | 98 | 99.8 | 100 | 100 | 0.2 | 99.7 | 99.5 - 99.8 | 100 | 94.1 | 99.5 | 100 | 100 | 0.5 | 99 | 98.4 - 99.6 | 100 | 96.7 | 99.5 | 100 | 100 | 0.5 | 99.3 | 99 - 99.7 |
| | PF – left | 100 | 95.6 | 100 | 100 | 100 | 0 | 98.6 | 97.3 - 99.8 | 100 | 93.2 | 99.5 | 100 | 100 | 0.5 | 98.1 | 96.5 - 99.7 | 100 | 95.8 | 99.5 | 100 | 100 | 0.5 | 98.3 | 96.9 - 99.7 |
| | PF – right | 100 | 97.6 | 100 | 100 | 100 | 0 | 99.1 | 97.9 - 100 | 100 | 96.4 | 99.6 | 100 | 100 | 0.4 | 98.8 | 97.6 - 100 | 100 | 97.6 | 99.6 | 100 | 100 | 0.4 | 98.9 | 97.8 - 100 |
| ***Structured in-lab overground 2MWT – smartphone vs Gait Up*** | | | | | | | | | | | | | | | | | | | | | | | | | |
| HC | BB | 100 | 96 | 98.7 | 100 | 100 | 1.3 | 98.9 | 98.4 - 99.5 | 98.7 | 95.6 | 97.4 | 99.4 | 100 | 2 | 98.3 | 97.8 - 98.8 | 98.7 | 96.6 | 98.1 | 99.4 | 100 | 1.3 | 98.6 | 98.2 - 99 |
| | BF | 99.8 | 96.3 | 98.7 | 100 | 100 | 1.3 | 98.9 | 98.3 - 99.5 | 98.1 | 94.7 | 96.4 | 99.4 | 100 | 3 | 97.8 | 97.2 - 98.4 | 98.7 | 95.9 | 97.5 | 99.6 | 100 | 2.1 | 98.3 | 97.8 - 98.9 |
| | PB – left | 98.1 | 91.9 | 95.8 | 99.5 | 100 | 3.6 | 97.3 | 96.4 - 98.3 | 98.8 | 96.2 | 98.2 | 99.9 | 100 | 1.7 | 98.6 | 98.1 - 99.2 | 98.5 | 94.9 | 97.5 | 99.2 | 99.7 | 1.7 | 97.9 | 97.4 - 98.5 |
| | PB – right | 97.9 | 92.7 | 95.7 | 99.5 | 100 | 3.8 | 97.2 | 96.3 - 98.2 | 99 | 95.5 | 97.5 | 99.5 | 100 | 2 | 98.6 | 98.1 - 99 | 98 | 95.5 | 96.9 | 99.2 | 99.7 | 2.3 | 97.9 | 97.3 - 98.4 |
| | PF – left | 97.9 | 91.8 | 94.8 | 99.4 | 100 | 4.7 | 96.9 | 95.9 - 97.9 | 98.3 | 95.3 | 96.7 | 99.5 | 100 | 2.9 | 98.1 | 97.5 - 98.7 | 97.8 | 94.4 | 96.5 | 98.5 | 99.7 | 1.9 | 97.5 | 96.9 - 98 |
| | PF – right | 98.1 | 92 | 94.8 | 99.5 | 100 | 4.7 | 97 | 96 - 98.1 | 98.4 | 93.6 | 96.5 | 99.3 | 100 | 2.8 | 97.7 | 96.9 - 98.5 | 97.6 | 94.7 | 96.4 | 98.9 | 99.5 | 2.5 | 97.3 | 96.6 - 98 |
| Mild PwMS | BB | 99.4 | 95.8 | 97.4 | 100 | 100 | 2.6 | 98.7 | 98 - 99.3 | 98.2 | 96.4 | 97.5 | 98.9 | 100 | 1.3 | 98.1 | 97.6 - 98.6 | 98.5 | 96.2 | 98 | 99.3 | 100 | 1.3 | 98.4 | 97.9 - 98.9 |
| | BF | 99.4 | 94.7 | 97.7 | 100 | 100 | 2.3 | 98.6 | 97.8 - 99.3 | 98 | 95 | 96 | 99.4 | 100 | 3.4 | 97.8 | 97 - 98.6 | 98.5 | 95 | 98 | 99.4 | 100 | 1.5 | 98.2 | 97.4 - 98.9 |
| | PB – left | 97.9 | 93 | 96 | 99.4 | 100 | 3.5 | 97.2 | 96 - 98.5 | 99.4 | 95.6 | 97.8 | 100 | 100 | 2.2 | 98.7 | 98.1 - 99.3 | 98.4 | 95.3 | 97.1 | 98.9 | 99.7 | 1.8 | 97.9 | 97.2 - 98.7 |
| | PB – right | 97.9 | 92.8 | 95.8 | 99.4 | 100 | 3.6 | 97.3 | 96.2 - 98.4 | 98.9 | 96.3 | 98.4 | 100 | 100 | 1.6 | 98.8 | 98.2 - 99.3 | 98.7 | 95.6 | 96.6 | 99.2 | 99.7 | 2.6 | 98 | 97.3 - 98.7 |
| | PF – left | 97.8 | 92.5 | 95.1 | 99.4 | 100 | 4.3 | 97.1 | 95.8 - 98.3 | 98.8 | 91.4 | 95.9 | 99.4 | 100 | 3.5 | 97.6 | 96.4 - 98.8 | 97.9 | 92 | 96.7 | 99.2 | 99.7 | 2.5 | 97.3 | 96.2 - 98.5 |
| | PF – right | 97.7 | 91.6 | 95.9 | 99.4 | 100 | 3.6 | 97 | 95.7 - 98.4 | 98.4 | 93.4 | 97.5 | 99.1 | 100 | 1.6 | 98 | 97.2 - 98.9 | 97.9 | 94.9 | 96.9 | 98.8 | 99.7 | 1.9 | 97.5 | 96.6 - 98.4 |
| Moderate PwMS | BB | 100 | 96.5 | 99.3 | 100 | 100 | 0.7 | 99.3 | 98.9 - 99.8 | 98.8 | 94.5 | 97.6 | 99.4 | 100 | 1.8 | 98.2 | 97.5 - 98.8 | 99.1 | 96.6 | 98.2 | 99.5 | 100 | 1.3 | 98.7 | 98.4 - 99.1 |
| | BF | 100 | 95.5 | 98.9 | 100 | 100 | 1.1 | 99.1 | 98.4 - 99.7 | 98.5 | 93.2 | 97 | 99.4 | 100 | 2.4 | 97.9 | 97.1 - 98.6 | 99.1 | 96 | 98 | 99.4 | 100 | 1.4 | 98.4 | 97.8 - 99.1 |
| | PB – left | 100 | 94.3 | 98.9 | 100 | 100 | 1.1 | 98.8 | 98 - 99.5 | 98.5 | 94.6 | 97.3 | 99.4 | 100 | 2.1 | 98 | 97.4 - 98.6 | 98.7 | 96.2 | 97.4 | 99.3 | 99.7 | 1.9 | 98.3 | 97.9 - 98.8 |
| | PB – right | 99.5 | 92.9 | 97.3 | 100 | 100 | 2.7 | 98.4 | 97.5 - 99.3 | 98.3 | 94.1 | 97.1 | 99.3 | 100 | 2.2 | 97.2 | 95.6 - 98.9 | 98.8 | 93.6 | 97.7 | 99.2 | 99.7 | 1.5 | 97.7 | 96.6 - 98.9 |
| | PF – left | 99.3 | 94.6 | 97.8 | 100 | 100 | 2.2 | 98.4 | 97.8 - 99.1 | 98.6 | 94.3 | 97.2 | 99.4 | 100 | 2.2 | 98.1 | 97.4 - 98.7 | 98.9 | 95.8 | 97.3 | 99.1 | 99.8 | 1.8 | 98.2 | 97.8 - 98.7 |
| | PF – right | 99.4 | 94.6 | 97.2 | 100 | 100 | 2.8 | 98.6 | 97.9 - 99.3 | 99 | 94.7 | 97.2 | 99.5 | 100 | 2.2 | 98.4 | 97.8 - 99 | 98.9 | 96.1 | 97.4 | 99.7 | 100 | 2.3 | 98.5 | 98 - 98.9 |
| Severe PwMS | BB | 100 | 93.3 | 98.4 | 100 | 100 | 1.6 | 96.3 | 91 - 100 | 98.7 | 94.5 | 98 | 99.3 | 100 | 1.4 | 95.8 | 90.5 - 100 | 99.2 | 94.1 | 98.4 | 99.6 | 99.8 | 1.2 | 96.1 | 90.7 - 100 |
| | BF | 100 | 97.2 | 98.5 | 100 | 100 | 1.5 | 98.2 | 95.9 - 100 | 99.2 | 96.2 | 98.2 | 100 | 100 | 1.8 | 97.9 | 95.9 - 99.8 | 99.3 | 97.2 | 98.2 | 99.7 | 100 | 1.5 | 98 | 95.9 - 100 |
| | PB – left | 99.3 | 96.4 | 98.1 | 100 | 100 | 1.9 | 98.8 | 98.2 - 99.4 | 98.5 | 94.8 | 97.4 | 99.3 | 100 | 1.9 | 97.9 | 97.2 - 98.6 | 98.7 | 95.6 | 97.6 | 99.4 | 100 | 1.8 | 98.3 | 97.8 - 98.9 |
| | PB – right | 99.3 | 95.8 | 98.4 | 100 | 100 | 1.6 | 98.4 | 97.1 - 99.7 | 97.9 | 91.3 | 96.9 | 98.7 | 100 | 1.9 | 97.1 | 95.9 - 98.3 | 98.4 | 94.8 | 97.3 | 99.3 | 99.8 | 2 | 97.7 | 96.6 - 98.8 |
| | PF – left | 99.3 | 95.2 | 98 | 100 | 100 | 2 | 98.3 | 97.3 - 99.4 | 98.8 | 92.5 | 97 | 99.4 | 100 | 2.4 | 97.1 | 95.1 - 99.1 | 98.7 | 95.5 | 97.4 | 99.3 | 100 | 2 | 97.7 | 96.2 - 99.2 |
| | PF – right | 99.3 | 88.7 | 98.1 | 100 | 100 | 1.9 | 97.5 | 95.6 - 99.4 | 97.8 | 87.6 | 96.9 | 98.6 | 100 | 1.7 | 96.5 | 94.5 - 98.5 | 98.2 | 87.2 | 97.7 | 99 | 99.8 | 1.3 | 97 | 95.1 - 98.9 |
| All PwMS | BB | 100 | 95.9 | 98.5 | 100 | 100 | 1.5 | 98.1 | 96.1 - 100 | 98.6 | 95 | 97.7 | 99.3 | 100 | 1.7 | 97.3 | 95.3 - 99.3 | 99 | 96.2 | 98.2 | 99.4 | 100 | 1.2 | 97.7 | 95.7 - 99.6 |
| | BF | 100 | 95.1 | 98.5 | 100 | 100 | 1.5 | 98.6 | 97.7 - 99.5 | 98.7 | 94.4 | 97.2 | 99.4 | 100 | 2.2 | 97.8 | 97 - 98.7 | 99.1 | 95.7 | 98 | 99.6 | 100 | 1.6 | 98.2 | 97.4 - 99.1 |
| | PB – left | 99.4 | 93.5 | 97.1 | 100 | 100 | 2.9 | 98.4 | 97.9 - 98.9 | 98.6 | 94.9 | 97.6 | 99.4 | 100 | 1.8 | 98.1 | 97.7 - 98.5 | 98.6 | 95.5 | 97.4 | 99.3 | 99.9 | 1.9 | 98.2 | 97.9 - 98.5 |
| | PB – right | 99.3 | 92.9 | 96.9 | 100 | 100 | 3.1 | 98.1 | 97.5 - 98.8 | 98.4 | 95.1 | 97.1 | 99.4 | 100 | 2.2 | 97.5 | 96.8 - 98.3 | 98.6 | 94.4 | 97.3 | 99.2 | 99.7 | 1.9 | 97.8 | 97.2 - 98.4 |
| | PF – left | 99.3 | 93.4 | 96.7 | 100 | 100 | 3.3 | 98.1 | 97.5 - 98.6 | 98.8 | 92.4 | 96.9 | 99.4 | 100 | 2.6 | 97.6 | 96.8 - 98.4 | 98.7 | 94.8 | 97.3 | 99.3 | 100 | 2 | 97.8 | 97.2 - 98.5 |
| | PF – right | 99.4 | 92.8 | 96.8 | 100 | 100 | 3.2 | 97.8 | 97 - 98.6 | 98.4 | 94 | 97.2 | 99.4 | 100 | 2.2 | 97.6 | 96.8 - 98.4 | 98.2 | 95.5 | 97.3 | 99.2 | 100 | 1.9 | 97.7 | 96.9 - 98.5 |
| ***Structured real-world walking – smartphone vs Gait Up*** | | | | | | | | | | | | | | | | | | | | | | | | | |
| HC | BF suggested location | 100 | 99.6 | 100 | 100 | 100 | 0 | 99.9 | 99.8 - 100 | 100 | 98.3 | 100 | 100 | 100 | 0.4 | 99.7 | 99.5 - 100 | 100 | 99 | 99.8 | 100 | 100 | 0.2 | 99.8 | 99.7 - 100 |
| Mild PwMS | BF suggested location | 100 | 99.4 | 100 | 100 | 100 | 0 | 99.9 | 99.8 - 100 | 100 | 96.8 | 100 | 100 | 100 | 0 | 99.1 | 97.7 - 100 | 100 | 97.5 | 100 | 100 | 100 | 0.2 | 99.2 | 97.9 - 100 |
| Moderate PwMS | BF suggested location | 100 | 98 | 99.7 | 100 | 100 | 0.3 | 99.5 | 98.9 - 100 | 99.5 | 92.3 | 97.3 | 100 | 100 | 1.8 | 98.3 | 97.3 - 99.2 | 99.7 | 95.5 | 98.5 | 100 | 100 | 1.2 | 98.8 | 98.1 - 99.5 |
| Severe PwMS | BF suggested location | 99.7 | 79.7 | 96.8 | 100 | 100 | 2.2 | 95.8 | 92.5 - 99.1 | 98.3 | 78.4 | 94 | 99.6 | 100 | 5.6 | 94.7 | 91.5 - 97.9 | 99 | 79.1 | 94.7 | 99.7 | 100 | 4 | 95.1 | 91.8 - 98.3 |
| All PwMS | BF suggested location | 100 | 91.3 | 99.3 | 100 | 100 | 0.7 | 98.2 | 96.8 - 99.5 | 99.5 | 87.4 | 97.4 | 100 | 100 | 2.3 | 97.1 | 95.7 - 98.5 | 99.6 | 87.3 | 98.2 | 100 | 100 | 1.7 | 97.5 | 96.1 - 98.8 |
| ***Unstructured real-world walking– smartphone vs Gait Up*** | | | | | | | | | | | | | | | | | | | | | | | | | |
| HC | User-preferred location | 97.5 | 92.8 | 95.5 | 98.5 | 99.3 | 4.8 | 96.7 | 95.8 - 97.6 | 98.8 | 92.9 | 97.1 | 99.3 | 99.6 | 3.8 | 97.7 | 96.9 - 98.5 | 97.8 | 92.9 | 95.8 | 98.6 | 99.5 | 4.7 | 97 | 96.2 - 97.8 |
| Mild PwMS | User-preferred location | 95.6 | 86.5 | 91.7 | 98 | 99.5 | 5.8 | 94.1 | 91.7 - 96.6 | 97.6 | 92 | 94.1 | 98.1 | 99.8 | 4.5 | 96.3 | 95.1 - 97.5 | 96.5 | 89.2 | 93.4 | 97.5 | 99.6 | 3.9 | 94.9 | 93.1 - 96.8 |
| Moderate PwMS | User-preferred location | 93.8 | 81.5 | 90.2 | 96 | 98.9 | 6.8 | 92.4 | 90.3 - 94.5 | 95.4 | 79.9 | 92.7 | 98 | 99.5 | 7.2 | 93.2 | 90.6 - 95.8 | 94.3 | 79.6 | 90.8 | 96.1 | 99.1 | 6.2 | 92.6 | 90.4 - 94.9 |
| Severe PwMS | User-preferred location | 90 | 75.2 | 86.5 | 94.2 | 97.9 | 9.9 | 89.1 | 86.6 - 91.6 | 92.6 | 78.1 | 87.5 | 97.2 | 98.7 | 11.8 | 91.3 | 88.9 - 93.7 | 90.9 | 79.4 | 85.5 | 94.6 | 97.2 | 9.7 | 89.8 | 87.7 - 91.9 |
| All PwMS | User-preferred location | 93.6 | 77.5 | 88.9 | 96.1 | 98.9 | 8.3 | 91.6 | 90.2 - 93 | 95.4 | 79.5 | 90.9 | 97.8 | 99.4 | 7.7 | 93.2 | 91.8 - 94.7 | 94 | 81.3 | 90 | 96.5 | 99.1 | 7 | 92.1 | 90.8 - 93.5 |

[a]Mild PwMS are PwMS with mild EDDS scores (EDSS: ≤ 3.5), moderate PwMS are PwMS with moderate EDSS scores (EDSS: 4.0 - 5.5), and severe PwMS are PwMS with severe EDSS scores (EDSS: 5.5 - 6.5).

BB = belt back; BF = belt front; CI = confidence interval; EDSS = Expanded Disability Status Scale; IQR = within-subject interquartile range; Mdn = median; P05 = 5th percentile; P95 = 95th percentile; PB = pocket back; PF = pocket front; PwMS = people with multiple sclerosis; Q1 = first quantile; Q3: = third quantile; ws-IQR = within-subject interquartile range.

**Supplementary Table S3. Algorithm error measures for initial contacts.**

| Cohort | Smartphone location | Constant error | | Absolute error | | Variable error | | Total variability | | Median error | | Median absolute error | | IQR error | |
|---|---|---|---|---|---|---|---|---|---|---|---|---|---|---|---|
| | | Mean | 95% CI | Mean | 95% CI | Mean | 95% CI | Mean | 95% CI | Mean | 95% CI | Mean | 95% CI | Mean | 95% CI |
| ***Structured in-lab treadmill 2MWT – smartphone vs mocap*** | | | | | | | | | | | | | | | |
| HC | Belt back | 0.06 | 0.05 - 0.06 | 0.06 | 0.05 - 0.06 | 0.01 | 0.01 - 0.02 | 0.06 | 0.05 - 0.06 | 0.06 | 0.05 - 0.06 | 0.06 | 0.05 - 0.06 | 0.02 | 0.02 - 0.02 |
| | Belt front | 0.04 | 0.03 - 0.05 | 0.04 | 0.03 - 0.05 | 0.02 | 0.01 - 0.02 | 0.04 | 0.04 - 0.05 | 0.04 | 0.03 - 0.05 | 0.04 | 0.03 - 0.05 | 0.02 | 0.02 - 0.03 |
| | Pocket back left | 0.08 | 0.07 - 0.09 | 0.08 | 0.07 - 0.09 | 0.03 | 0.03 - 0.03 | 0.08 | 0.08 - 0.09 | 0.08 | 0.07 - 0.09 | 0.08 | 0.07 - 0.09 | 0.05 | 0.04 - 0.06 |
| | Pocket back right | 0.08 | 0.07 - 0.08 | 0.08 | 0.07 - 0.08 | 0.03 | 0.03 - 0.04 | 0.08 | 0.08 - 0.09 | 0.07 | 0.07 - 0.08 | 0.07 | 0.07 - 0.08 | 0.06 | 0.06 - 0.07 |
| | Pocket front left | 0.05 | 0.04 - 0.05 | 0.05 | 0.04 - 0.06 | 0.04 | 0.03 - 0.04 | 0.06 | 0.05 - 0.07 | 0.05 | 0.04 - 0.05 | 0.05 | 0.04 - 0.05 | 0.06 | 0.06 - 0.07 |
| | Pocket front right | 0.04 | 0.04 - 0.05 | 0.04 | 0.04 - 0.05 | 0.03 | 0.03 - 0.03 | 0.05 | 0.05 - 0.06 | 0.04 | 0.04 - 0.05 | 0.04 | 0.04 - 0.05 | 0.05 | 0.05 - 0.06 |
| All PwMS | Belt back | 0.07 | 0.06 - 0.08 | 0.07 | 0.07 - 0.08 | 0.02 | 0.02 - 0.02 | 0.08 | 0.07 - 0.08 | 0.07 | 0.06 - 0.08 | 0.07 | 0.06 - 0.08 | 0.03 | 0.02 - 0.03 |
| | Belt front | 0.05 | 0.04 - 0.06 | 0.05 | 0.05 - 0.06 | 0.02 | 0.02 - 0.03 | 0.06 | 0.05 - 0.07 | 0.05 | 0.04 - 0.06 | 0.05 | 0.04 - 0.06 | 0.03 | 0.03 - 0.03 |
| | Pocket back left | 0.07 | 0.06 - 0.09 | 0.08 | 0.08 - 0.09 | 0.03 | 0.03 - 0.03 | 0.09 | 0.08 - 0.1 | 0.07 | 0.06 - 0.08 | 0.08 | 0.07 - 0.09 | 0.05 | 0.04 - 0.05 |
| | Pocket back right | 0.08 | 0.07 - 0.08 | 0.08 | 0.07 - 0.09 | 0.04 | 0.03 - 0.04 | 0.09 | 0.08 - 0.09 | 0.08 | 0.07 - 0.08 | 0.08 | 0.07 - 0.08 | 0.05 | 0.05 - 0.06 |
| | Pocket front left | 0.06 | 0.05 - 0.07 | 0.07 | 0.06 - 0.08 | 0.04 | 0.03 - 0.04 | 0.08 | 0.07 - 0.08 | 0.06 | 0.05 - 0.07 | 0.06 | 0.06 - 0.07 | 0.06 | 0.05 - 0.07 |
| | Pocket front right | 0.06 | 0.05 - 0.06 | 0.06 | 0.05 - 0.07 | 0.03 | 0.03 - 0.04 | 0.07 | 0.06 - 0.07 | 0.05 | 0.05 - 0.06 | 0.06 | 0.05 - 0.06 | 0.05 | 0.05 - 0.06 |
| ***Structured in-lab treadmill 2MWT – smartphone vs Gait Up*** | | | | | | | | | | | | | | | |
| HC | Belt back | 0.04 | 0.03 - 0.05 | 0.04 | 0.03 - 0.05 | 0.02 | 0.01 - 0.02 | 0.04 | 0.04 - 0.05 | 0.04 | 0.03 - 0.05 | 0.04 | 0.03 - 0.05 | 0.02 | 0.02 - 0.02 |
| | Belt front | 0.02 | 0.01 - 0.03 | 0.03 | 0.02 - 0.03 | 0.02 | 0.02 - 0.02 | 0.03 | 0.02 - 0.04 | 0.02 | 0.01 - 0.03 | 0.03 | 0.02 - 0.03 | 0.03 | 0.02 - 0.03 |
| | Pocket back left | 0.06 | 0.06 - 0.07 | 0.06 | 0.06 - 0.07 | 0.03 | 0.03 - 0.03 | 0.07 | 0.06 - 0.08 | 0.06 | 0.06 - 0.07 | 0.06 | 0.06 - 0.07 | 0.05 | 0.04 - 0.06 |
| | Pocket back right | 0.06 | 0.05 - 0.07 | 0.06 | 0.06 - 0.07 | 0.04 | 0.03 - 0.04 | 0.07 | 0.07 - 0.08 | 0.06 | 0.05 - 0.07 | 0.06 | 0.05 - 0.07 | 0.07 | 0.06 - 0.07 |

| | Pocket front left | 0.03 | 0.02 - 0.04 | 0.04 | 0.04 - 0.05 | 0.04 | 0.03 - 0.04 | 0.05 | 0.04 - 0.05 | 0.03 | 0.02 - 0.03 | 0.04 | 0.03 - 0.04 | 0.06 | 0.06 - 0.07 |
|---|---|---|---|---|---|---|---|---|---|---|---|---|---|---|---|
| | Pocket front right | 0.03 | 0.02 - 0.03 | 0.04 | 0.03 - 0.04 | 0.03 | 0.03 - 0.03 | 0.04 | 0.04 - 0.05 | 0.02 | 0.02 - 0.03 | 0.03 | 0.03 - 0.04 | 0.06 | 0.05 - 0.06 |
| All PwMS | Belt back | 0.05 | 0.05 - 0.06 | 0.06 | 0.05 - 0.06 | 0.02 | 0.02 - 0.03 | 0.06 | 0.05 - 0.07 | 0.05 | 0.05 - 0.06 | 0.05 | 0.05 - 0.06 | 0.03 | 0.03 - 0.03 |
| | Belt front | 0.03 | 0.02 - 0.04 | 0.04 | 0.04 - 0.05 | 0.03 | 0.02 - 0.03 | 0.05 | 0.04 - 0.06 | 0.03 | 0.02 - 0.04 | 0.04 | 0.03 - 0.05 | 0.03 | 0.03 - 0.04 |
| | Pocket back left | 0.06 | 0.04 - 0.07 | 0.07 | 0.06 - 0.08 | 0.03 | 0.03 - 0.04 | 0.08 | 0.07 - 0.09 | 0.06 | 0.04 - 0.07 | 0.07 | 0.06 - 0.08 | 0.05 | 0.04 - 0.05 |
| | Pocket back right | 0.06 | 0.06 - 0.07 | 0.07 | 0.06 - 0.07 | 0.04 | 0.03 - 0.04 | 0.07 | 0.07 - 0.08 | 0.06 | 0.05 - 0.07 | 0.06 | 0.06 - 0.07 | 0.06 | 0.05 - 0.06 |
| | Pocket front left | 0.05 | 0.04 - 0.06 | 0.06 | 0.05 - 0.07 | 0.04 | 0.03 - 0.04 | 0.07 | 0.06 - 0.07 | 0.05 | 0.04 - 0.06 | 0.05 | 0.05 - 0.06 | 0.06 | 0.05 - 0.07 |
| | Pocket front right | 0.04 | 0.03 - 0.05 | 0.05 | 0.04 - 0.06 | 0.03 | 0.03 - 0.04 | 0.06 | 0.05 - 0.06 | 0.04 | 0.03 - 0.05 | 0.05 | 0.04 - 0.05 | 0.05 | 0.05 - 0.06 |
| ***Structured in-lab overground 2MWT – smartphone vs Gait Up*** | | | | | | | | | | | | | | | |
| HC | Belt back | 0.03 | 0.03 - 0.04 | 0.04 | 0.03 - 0.04 | 0.02 | 0.02 - 0.02 | 0.04 | 0.03 - 0.04 | 0.03 | 0.03 - 0.04 | 0.03 | 0.03 - 0.04 | 0.03 | 0.02 - 0.03 |
| | Belt front | 0.02 | 0.01 - 0.02 | 0.03 | 0.02 - 0.03 | 0.02 | 0.02 - 0.03 | 0.03 | 0.03 - 0.04 | 0.02 | 0.01 - 0.02 | 0.02 | 0.02 - 0.03 | 0.03 | 0.02 - 0.04 |
| | Pocket back left | 0.06 | 0.05 - 0.06 | 0.06 | 0.05 - 0.06 | 0.03 | 0.03 - 0.04 | 0.07 | 0.06 - 0.07 | 0.06 | 0.05 - 0.07 | 0.06 | 0.05 - 0.07 | 0.05 | 0.05 - 0.06 |
| | Pocket back right | 0.05 | 0.04 - 0.06 | 0.06 | 0.05 - 0.06 | 0.04 | 0.04 - 0.04 | 0.07 | 0.06 - 0.07 | 0.05 | 0.04 - 0.06 | 0.06 | 0.05 - 0.06 | 0.07 | 0.06 - 0.08 |
| | Pocket front left | 0.03 | 0.03 - 0.04 | 0.05 | 0.05 - 0.05 | 0.05 | 0.04 - 0.05 | 0.06 | 0.05 - 0.06 | 0.03 | 0.02 - 0.04 | 0.05 | 0.04 - 0.05 | 0.08 | 0.08 - 0.09 |
| | Pocket front right | 0.02 | 0.02 - 0.03 | 0.04 | 0.04 - 0.04 | 0.04 | 0.04 - 0.04 | 0.05 | 0.04 - 0.05 | 0.02 | 0.01 - 0.03 | 0.03 | 0.03 - 0.04 | 0.07 | 0.06 - 0.07 |
| All PwMS | Belt back | 0.03 | 0.03 - 0.04 | 0.04 | 0.04 - 0.04 | 0.03 | 0.02 - 0.03 | 0.05 | 0.04 - 0.05 | 0.03 | 0.03 - 0.04 | 0.04 | 0.03 - 0.04 | 0.04 | 0.03 - 0.04 |
| | Belt front | 0.02 | 0.02 - 0.03 | 0.03 | 0.03 - 0.04 | 0.03 | 0.02 - 0.03 | 0.04 | 0.04 - 0.04 | 0.02 | 0.02 - 0.03 | 0.03 | 0.03 - 0.04 | 0.04 | 0.03 - 0.04 |
| | Pocket back left | 0.05 | 0.04 - 0.05 | 0.06 | 0.05 - 0.06 | 0.04 | 0.04 - 0.04 | 0.07 | 0.06 - 0.07 | 0.05 | 0.04 - 0.05 | 0.05 | 0.05 - 0.06 | 0.06 | 0.05 - 0.06 |
| | Pocket back right | 0.05 | 0.05 - 0.06 | 0.06 | 0.05 - 0.06 | 0.04 | 0.04 - 0.05 | 0.07 | 0.07 - 0.07 | 0.05 | 0.04 - 0.05 | 0.05 | 0.05 - 0.06 | 0.07 | 0.06 - 0.07 |
| | Pocket front left | 0.04 | 0.03 - 0.04 | 0.05 | 0.05 - 0.05 | 0.05 | 0.04 - 0.05 | 0.06 | 0.06 - 0.07 | 0.04 | 0.03 - 0.04 | 0.05 | 0.04 - 0.05 | 0.07 | 0.07 - 0.08 |
| | Pocket front right | 0.03 | 0.02 - 0.03 | 0.04 | 0.04 - 0.05 | 0.04 | 0.04 - 0.04 | 0.05 | 0.05 - 0.06 | 0.03 | 0.02 - 0.03 | 0.04 | 0.03 - 0.04 | 0.06 | 0.06 - 0.07 |
| ***Structured real-world walking – smartphone vs Gait Up*** | | | | | | | | | | | | | | | |
| HC | Belt front suggested | 0.05 | 0.05 - 0.06 | 0.06 | 0.05 - 0.06 | 0.02 | 0.02 - 0.03 | 0.06 | 0.06 - 0.07 | 0.05 | 0.05 - 0.06 | 0.05 | 0.05 - 0.06 | 0.03 | 0.03 - 0.04 |
| All PwMS | Belt front suggested | 0.06 | 0.05 - 0.06 | 0.07 | 0.06 - 0.07 | 0.04 | 0.03 - 0.04 | 0.07 | 0.07 - 0.08 | 0.06 | 0.05 - 0.06 | 0.06 | 0.06 - 0.07 | 0.05 | 0.04 - 0.05 |
| ***Unstructured real-world walking– smartphone vs Gait Up*** | | | | | | | | | | | | | | | |
| HC | User-selected location | 0.06 | 0.05 - 0.06 | 0.07 | 0.06 - 0.08 | 0.05 | 0.04 - 0.05 | 0.08 | 0.07 - 0.09 | 0.06 | 0.05 - 0.06 | 0.07 | 0.06 - 0.07 | 0.06 | 0.05 - 0.08 |
| All PwMS | User-selected location | 0.07 | 0.06 - 0.07 | 0.08 | 0.08 - 0.09 | 0.06 | 0.06 - 0.06 | 0.09 | 0.09 - 0.1 | 0.07 | 0.06 - 0.07 | 0.08 | 0.07 - 0.08 | 0.07 | 0.07 - 0.08 |

BB = belt back; BF = belt front; CI = confidence interval; IQR = interquartile range; PB = pocket back; PF = pocket front; PwMS = people with multiple sclerosis.

**Supplementary Table S4. Algorithm error measures for final contacts.**

| Cohort | Smartphone location | Constant error | | Absolute error | | Variable error | | Total variability | | Median error | | Median absolute error | | IQR error | |
|---|---|---|---|---|---|---|---|---|---|---|---|---|---|---|---|
| | | Mean | 95% CI | Mean | 95% CI | Mean | 95% CI | Mean | 95% CI | Mean | 95% CI | Mean | 95% CI | Mean | 95% CI |
| ***Structured in-lab treadmill 2MWT – smartphone vs mocap*** | | | | | | | | | | | | | | | |
| HC | Belt back | 0.03 | 0.03 - 0.04 | 0.03 | 0.03 - 0.04 | 0.02 | 0.01 - 0.02 | 0.04 | 0.03 - 0.04 | 0.03 | 0.03 - 0.04 | 0.03 | 0.03 - 0.04 | 0.02 | 0.02 - 0.02 |
| | Belt front | 0.02 | 0.01 - 0.03 | 0.03 | 0.02 - 0.03 | 0.02 | 0.02 - 0.02 | 0.03 | 0.03 - 0.04 | 0.02 | 0.01 - 0.03 | 0.03 | 0.02 - 0.03 | 0.03 | 0.02 - 0.03 |
| | Pocket back left | 0.04 | 0.04 - 0.05 | 0.04 | 0.04 - 0.05 | 0.02 | 0.02 - 0.02 | 0.05 | 0.04 - 0.05 | 0.04 | 0.04 - 0.05 | 0.04 | 0.04 - 0.05 | 0.03 | 0.02 - 0.03 |
| | Pocket back right | 0.04 | 0.03 - 0.05 | 0.04 | 0.04 - 0.05 | 0.03 | 0.02 - 0.03 | 0.05 | 0.04 - 0.05 | 0.04 | 0.04 - 0.05 | 0.05 | 0.04 - 0.05 | 0.04 | 0.04 - 0.05 |
| | Pocket front left | 0.02 | 0.01 - 0.02 | 0.04 | 0.04 - 0.05 | 0.04 | 0.04 - 0.05 | 0.05 | 0.04 - 0.05 | 0.03 | 0.02 - 0.03 | 0.04 | 0.04 - 0.05 | 0.08 | 0.07 - 0.08 |
| | Pocket front right | 0.01 | 0.01 - 0.02 | 0.04 | 0.03 - 0.04 | 0.04 | 0.03 - 0.04 | 0.04 | 0.04 - 0.05 | 0.02 | 0.01 - 0.03 | 0.04 | 0.03 - 0.04 | 0.07 | 0.06 - 0.07 |
| All PwMS | Belt back | 0.01 | 0.0 - 0.02 | 0.04 | 0.04 - 0.04 | 0.03 | 0.02 - 0.03 | 0.04 | 0.04 - 0.05 | 0.01 | 0.0 - 0.02 | 0.04 | 0.03 - 0.04 | 0.03 | 0.03 - 0.04 |
| | Belt front | -0.01 | -0.02 to -0.0 | 0.04 | 0.04 - 0.05 | 0.03 | 0.03 - 0.03 | 0.05 | 0.05 - 0.06 | -0.01 | -0.02 to -0.0 | 0.04 | 0.04 - 0.05 | 0.04 | 0.04 - 0.04 |
| | Pocket back left | 0.02 | 0.01 - 0.03 | 0.05 | 0.04 - 0.05 | 0.03 | 0.03 - 0.03 | 0.05 | 0.05 - 0.06 | 0.02 | 0.01 - 0.03 | 0.05 | 0.04 - 0.05 | 0.04 | 0.04 - 0.05 |
| | Pocket back right | 0.01 | 0.0 - 0.02 | 0.05 | 0.04 - 0.05 | 0.04 | 0.03 - 0.04 | 0.05 | 0.05 - 0.06 | 0.01 | 0.01 - 0.02 | 0.05 | 0.04 - 0.05 | 0.06 | 0.05 - 0.06 |
| | Pocket front left | 0.01 | 0.0 - 0.02 | 0.05 | 0.04 - 0.06 | 0.05 | 0.04 - 0.05 | 0.06 | 0.05 - 0.06 | 0.02 | 0.01 - 0.03 | 0.05 | 0.04 - 0.05 | 0.07 | 0.07 - 0.08 |
| | Pocket front right | 0 | -0.01 to 0.0 | 0.05 | 0.04 - 0.05 | 0.04 | 0.04 - 0.05 | 0.05 | 0.05 - 0.06 | 0 | -0.01 - 0.01 | 0.04 | 0.04 - 0.05 | 0.07 | 0.06 - 0.08 |
| ***Structured in-lab treadmill 2MWT – smartphone vs Gait Up*** | | | | | | | | | | | | | | | |
| HC | Belt back | 0.06 | 0.06 - 0.07 | 0.06 | 0.06 - 0.07 | 0.02 | 0.02 - 0.02 | 0.07 | 0.06 - 0.07 | 0.06 | 0.06 - 0.07 | 0.06 | 0.06 - 0.07 | 0.02 | 0.02 - 0.03 |
| | Belt front | 0.05 | 0.04 - 0.06 | 0.05 | 0.04 - 0.06 | 0.02 | 0.02 - 0.02 | 0.06 | 0.05 - 0.06 | 0.05 | 0.04 - 0.06 | 0.05 | 0.04 - 0.06 | 0.03 | 0.02 - 0.03 |
| | Pocket back left | 0.07 | 0.07 - 0.08 | 0.07 | 0.07 - 0.08 | 0.02 | 0.02 - 0.02 | 0.08 | 0.07 - 0.08 | 0.07 | 0.07 - 0.08 | 0.07 | 0.07 - 0.08 | 0.03 | 0.03 - 0.04 |
| | Pocket back right | 0.07 | 0.07 - 0.08 | 0.07 | 0.07 - 0.08 | 0.03 | 0.02 - 0.03 | 0.08 | 0.07 - 0.08 | 0.08 | 0.07 - 0.08 | 0.08 | 0.07 - 0.08 | 0.04 | 0.04 - 0.05 |
| | Pocket front left | 0.05 | 0.04 - 0.05 | 0.06 | 0.05 - 0.06 | 0.04 | 0.04 - 0.05 | 0.07 | 0.06 - 0.07 | 0.06 | 0.05 - 0.06 | 0.06 | 0.06 - 0.07 | 0.08 | 0.07 - 0.09 |
| | Pocket front right | 0.04 | 0.04 - 0.05 | 0.05 | 0.05 - 0.06 | 0.04 | 0.04 - 0.04 | 0.06 | 0.06 - 0.07 | 0.05 | 0.04 - 0.06 | 0.06 | 0.05 - 0.06 | 0.07 | 0.06 - 0.08 |
| All PwMS | Belt back | 0.04 | 0.03 - 0.05 | 0.06 | 0.05 - 0.07 | 0.03 | 0.03 - 0.03 | 0.07 | 0.06 - 0.07 | 0.04 | 0.03 - 0.05 | 0.06 | 0.05 - 0.07 | 0.04 | 0.04 - 0.04 |
| | Belt front | 0.01 | 0.0 - 0.03 | 0.06 | 0.05 - 0.06 | 0.03 | 0.03 - 0.04 | 0.06 | 0.06 - 0.07 | 0.01 | 0.0 - 0.03 | 0.05 | 0.05 - 0.06 | 0.04 | 0.04 - 0.05 |
| | Pocket back left | 0.04 | 0.02 - 0.05 | 0.07 | 0.06 - 0.08 | 0.03 | 0.03 - 0.04 | 0.07 | 0.07 - 0.08 | 0.04 | 0.03 - 0.06 | 0.07 | 0.06 - 0.08 | 0.05 | 0.04 - 0.05 |
| | Pocket back right | 0.04 | 0.03 - 0.05 | 0.06 | 0.06 - 0.07 | 0.04 | 0.03 - 0.04 | 0.07 | 0.06 - 0.08 | 0.05 | 0.03 - 0.06 | 0.06 | 0.05 - 0.07 | 0.06 | 0.05 - 0.06 |
| | Pocket front left | 0.04 | 0.03 - 0.05 | 0.06 | 0.05 - 0.07 | 0.05 | 0.04 - 0.05 | 0.07 | 0.06 - 0.08 | 0.05 | 0.04 - 0.06 | 0.06 | 0.05 - 0.07 | 0.08 | 0.07 - 0.08 |
| | Pocket front right | 0.03 | 0.02 - 0.03 | 0.05 | 0.05 - 0.06 | 0.04 | 0.04 - 0.05 | 0.06 | 0.06 - 0.07 | 0.03 | 0.02 - 0.04 | 0.05 | 0.05 - 0.06 | 0.07 | 0.06 - 0.07 |
| ***Structured in-lab overground 2MWT – smartphone vs Gait Up*** | | | | | | | | | | | | | | | |
| HC | Belt back | 0.06 | 0.06 - 0.07 | 0.06 | 0.06 - 0.07 | 0.02 | 0.02 - 0.02 | 0.07 | 0.06 - 0.07 | 0.06 | 0.06 - 0.07 | 0.06 | 0.06 - 0.07 | 0.03 | 0.02 - 0.03 |
| | Belt front | 0.06 | 0.05 - 0.07 | 0.06 | 0.05 - 0.07 | 0.02 | 0.02 - 0.03 | 0.06 | 0.06 - 0.07 | 0.06 | 0.05 - 0.07 | 0.06 | 0.05 - 0.07 | 0.03 | 0.02 - 0.03 |
| | Pocket back left | 0.07 | 0.07 - 0.08 | 0.07 | 0.07 - 0.08 | 0.03 | 0.02 - 0.03 | 0.08 | 0.07 - 0.09 | 0.07 | 0.07 - 0.08 | 0.07 | 0.07 - 0.08 | 0.04 | 0.03 - 0.04 |
| | Pocket back right | 0.07 | 0.06 - 0.08 | 0.07 | 0.06 - 0.08 | 0.03 | 0.03 - 0.03 | 0.08 | 0.07 - 0.08 | 0.07 | 0.06 - 0.08 | 0.07 | 0.06 - 0.08 | 0.05 | 0.04 - 0.05 |
| | Pocket front left | 0.06 | 0.05 - 0.06 | 0.06 | 0.06 - 0.07 | 0.04 | 0.04 - 0.05 | 0.07 | 0.07 - 0.08 | 0.06 | 0.05 - 0.07 | 0.06 | 0.06 - 0.07 | 0.08 | 0.07 - 0.08 |
| | Pocket front right | 0.05 | 0.04 - 0.06 | 0.05 | 0.05 - 0.06 | 0.04 | 0.04 - 0.04 | 0.06 | 0.06 - 0.07 | 0.05 | 0.04 - 0.06 | 0.05 | 0.05 - 0.06 | 0.07 | 0.06 - 0.08 |
| All PwMS | Belt back | 0.03 | 0.03 - 0.04 | 0.05 | 0.04 - 0.05 | 0.03 | 0.03 - 0.03 | 0.06 | 0.05 - 0.06 | 0.03 | 0.03 - 0.04 | 0.05 | 0.04 - 0.05 | 0.04 | 0.03 - 0.04 |
| | Belt front | 0.02 | 0.01 - 0.03 | 0.05 | 0.05 - 0.05 | 0.03 | 0.03 - 0.04 | 0.06 | 0.05 - 0.06 | 0.02 | 0.01 - 0.03 | 0.05 | 0.04 - 0.05 | 0.05 | 0.04 - 0.05 |
| | Pocket back left | 0.04 | 0.03 - 0.05 | 0.06 | 0.05 - 0.06 | 0.04 | 0.03 - 0.04 | 0.06 | 0.06 - 0.07 | 0.04 | 0.03 - 0.05 | 0.05 | 0.05 - 0.06 | 0.05 | 0.04 - 0.05 |
| | Pocket back right | 0.04 | 0.03 - 0.05 | 0.06 | 0.05 - 0.06 | 0.04 | 0.03 - 0.04 | 0.07 | 0.06 - 0.07 | 0.04 | 0.03 - 0.05 | 0.05 | 0.05 - 0.06 | 0.05 | 0.05 - 0.06 |
| | Pocket front left | 0.03 | 0.03 - 0.04 | 0.06 | 0.06 - 0.06 | 0.05 | 0.05 - 0.06 | 0.07 | 0.06 - 0.07 | 0.03 | 0.03 - 0.04 | 0.06 | 0.05 - 0.06 | 0.09 | 0.08 - 0.1 |
| | Pocket front right | 0.03 | 0.02 - 0.03 | 0.05 | 0.05 - 0.05 | 0.04 | 0.04 - 0.05 | 0.06 | 0.05 - 0.06 | 0.03 | 0.02 - 0.04 | 0.05 | 0.04 - 0.05 | 0.07 | 0.06 - 0.07 |
| ***Structured real-world walking – smartphone vs Gait Up*** | | | | | | | | | | | | | | | |
| HC | Belt front suggested | 0.08 | 0.08 - 0.09 | 0.08 | 0.08 - 0.09 | 0.03 | 0.02 - 0.03 | 0.09 | 0.08 - 0.1 | 0.08 | 0.08 - 0.09 | 0.08 | 0.08 - 0.09 | 0.03 | 0.03 - 0.04 |
| All PwMS | Belt front suggested | 0.05 | 0.04 - 0.06 | 0.08 | 0.07 - 0.08 | 0.04 | 0.04 - 0.05 | 0.08 | 0.08 - 0.09 | 0.05 | 0.04 - 0.07 | 0.07 | 0.07 - 0.08 | 0.06 | 0.05 - 0.06 |
| ***Unstructured real-world walking– smartphone vs Gait Up*** | | | | | | | | | | | | | | | |
| HC | User-selected location | 0.07 | 0.06 - 0.07 | 0.08 | 0.08 - 0.09 | 0.06 | 0.05 - 0.06 | 0.09 | 0.09 - 0.1 | 0.07 | 0.06 - 0.08 | 0.08 | 0.07 - 0.09 | 0.07 | 0.06 - 0.08 |
| All PwMS | User-selected location | 0.05 | 0.05 - 0.06 | 0.08 | 0.08 - 0.09 | 0.07 | 0.06 - 0.07 | 0.1 | 0.09 - 0.1 | 0.06 | 0.05 - 0.07 | 0.08 | 0.08 - 0.09 | 0.09 | 0.08 - 0.09 |

BB = belt back; BF = belt front; CI = confidence interval; IQR = interquartile range; PB = pocket back; PF = pocket front; PwMS = people with multiple sclerosis.

**Supplementary Table S5. Parameter estimates of the Bayesian model examining the influence of different factors on the F1 score for initial contacts.**

| Parameter | Mean | Median | Std | 5% | 95% | IQR | Z-score | P > \|z\| |
|---|---|---|---|---|---|---|---|---|
| Female - Male | -0.197 | -0.204 | 1.393 | -2.421 | 2.076 | 1.891 | -0.142 | 0.887 |
| Indoors - Outdoors | -1.788 | -1.779 | 1.359 | -4.021 | 0.463 | 1.839 | -1.315 | 0.188 |
| With walking aid vs without walking aid | -0.697 | -0.719 | 1.409 | -3.047 | 1.582 | 1.877 | -0.495 | 0.621 |
| Disease | -1.769 | -1.77 | 1.01 | -3.42 | -0.123 | 1.373 | -1.751 | 0.08 |

**Supplementary Table S6. Parameter estimates of the Bayesian model examining the influence of different factors on the F1 score for final contacts.**

| Parameter | Mean | Median | Std | 5% | 95% | IQR | Z-score | P > \|z\| |
|---|---|---|---|---|---|---|---|---|
| Female vs male | -0.168 | -0.174 | 1.388 | -2.437 | 2.135 | 1.861 | -0.121 | 0.904 |
| Indoors vs outdoors | -2.061 | -2.052 | 1.356 | -4.308 | 0.169 | 1.834 | -1.52 | 0.129 |
| With walking aid vs without walking aid | -0.426 | -0.421 | 1.419 | -2.738 | 1.937 | 1.847 | -0.3 | 0.764 |
| Disease | -1.784 | -1.791 | 1.0 | -3.401 | -0.171 | 1.369 | -1.783 | 0.075 |